\documentclass{cernyrep} 
\usepackage{texnames}
\usepackage[T1]{fontenc}
\usepackage{bbm}
\newcommand{\be}{\begin{equation}}
\newcommand{\ee}{\end{equation}}
\newcommand{\bi}{\begin{itemize}}
\newcommand{\ei}{\end{itemize}}
\newcommand{\nn}{\nonumber}

\begin{document}
\title{\boldmath Introduction to Flavour Physics and $CP$ Violation}
 
\author{Monika Blanke}

\institute{{Institut fur Kernphysik, Karlsruhe Institute of Technology,
  Hermann-von-Helmholtz-Platz 1,\\
  D-76344 Eggenstein-Leopoldshafen, Germany}\\
 {Institut fur Theoretische Teilchenphysik,
  Karlsruhe Institute of Technology, Engesserstra\ss e 7,\\
  D-76128 Karlsruhe, Germany}}

\maketitle 

\begin{abstract}
These lecture notes provide an introduction to the theoretical concepts of flavour physics and $CP$ violation, based on a series of lectures given at the ESHEP 2016 summer school. In the first lecture we review the basics of flavour and $CP$ violation in the Standard Model. The second lecture is dedicated to the phenomenology of $K$ and $B$ meson decays, where we focus on a few representative observables. In the third lecture we give an introduction to flavour physics beyond the Standard Model, both within the framework of Minimal Flavour Violation and beyond.
\end{abstract}

\section*{Introduction}

Flavour physics and $CP$ violation have played a central role in the development of the Standard Model (SM). It was the underlying $SU(3)$ flavour symmetry of mesons and baryons that lead Gell-Mann to the introduction of up, down, and strange quarks as the fundamental constituents of hadronic matter \cite{GellMann:1964nj}. The charm quark was predicted prior to its discovery as an explanation for the smallness of the $K_L\to\mu^+\mu^-$ decay rate \cite{Glashow:1970gm}. It was then realized that, in order to account for the observed $CP$ violation in neutral kaon mixing, a third generation of quarks was needed \cite{Kobayashi:1973fv}. Furthermore, the heaviness of the top quark was predicted from the size of $CP$ violation in $K^0-\bar K^0$ mixing and of the neutral $B$ meson oscillations prior to its discovery \cite{Buras:1993wr}.

Subsequently, the role of flavour physics has shifted from the discovery of the building blocks of the SM to the measurement of its parameters. The majority of the SM parameters is related to the flavour sector and can thus be determined in flavour violating decays. With increasing experimental and theoretical accuracy, their determination has by now reached an impressive precision. 

Having at hand a good understanding of the SM flavour sector, the measurement of flavour and $CP$-violating processes can be used to put constraints on models of New Physics (NP). Due to the strong suppression of flavour violation in the SM, very high energy scales can be probed in this way, well beyond the reach of direct searches for new particles in high energy collisions at the Large Hadron Collider (LHC) (see \cite{Buras:2013ooa} for a recent review). NP around the TeV scale is therefore required to have a highly non-trivial flavour structure.

The present lecture notes provide a summary of a series of three lectures given at the European School for High Energy Physics (ESHEP) 2016 in Skeikampen, Norway. The topic of these lectures is restricted to quark flavour physics. Flavour in the charged lepton and neutrino sector has been covered by Gabriela Barenboim \cite{Barenboim:2016ili}.
In lecture \ref{sec:SM} we introduce the quark Yukawa couplings and the Cabibbo-Kobayashi-Maskawa (CKM) matrix as the basic ingredients of flavour and $CP$ violation in the SM. We then discuss the physics of flavour changing neutral currents and review their description in terms of an effective Hamiltonian and the path from quark level flavour transitions to the decays of mesons. Lecture \ref{sec:pheno} is devoted to the phenomenology of flavour and $CP$-violating decays of kaons and $B$ mesons. Rather than providing an exhaustive overview, we focus on a number of particularly interesting benchmark processes, like neutral meson mixings, rare $K$ meson decays, and the recently observed anomalies in $B$ meson decays based on semileptonic $b\to s$ transitions. In lecture \ref{sec:NP} we turn our attention to flavour physics beyond the SM. After reviewing the generic constraints on the scale of NP from flavour and $CP$-violating decays, we discuss two well-known suppression mechanisms for the size of new flavour violating interactions. We motivate the concept of Minimal Flavour Violation (MFV) from the flavour symmetries of the SM. As an alternative to MFV, we also discuss how models with partially composite fermions can explain the observed flavour hierarchies in the SM, and at the same time suppress flavour changing neutral currents to an acceptable level.

A large number of excellent lecture notes on the physics of flavour and $CP$ violation can be found on the arXiv. As a few representative examples, let me recommend especially the lectures by Gino Isidori \cite{Isidori:2013ez}, Yuval Grossmann \cite{Grossman:2010gw}, and Andrzej j.\ Buras \cite{Buras:2005xt}. An extensive pedagogical introduction into the technicalities of the theory of flavour physics can be found in \cite{Buras:1998raa}.

\section{Flavour Physics in the Standard Model}\label{sec:SM}

\subsection{Quark Yukawa Couplings and the CKM Matrix}

In nature, all fundamental matter fields -- quarks, charged leptons, and neutrinos -- come in three copies, the so-called {\it flavours}. They can be collected in three fermion generations, with increasing masses, but otherwise identical quantum numbers. The subject of flavour physics is the description of interactions between the various flavours, with the goal to unravel the underlying dynamics of flavour symmetry breaking. 

In the SM, the left-handed quarks are arranged in doublets of the $SU(2)_L$ weak interactions:
\be\label{eq:quarks}
Q_j =
\begin{pmatrix}
u_L \\ d_L
\end{pmatrix},
\begin{pmatrix}
c_L \\ s_L
\end{pmatrix},
\begin{pmatrix}
t_L \\ b_L
\end{pmatrix}\,,
\ee
while the right-handed quarks are introduced as $SU(2)_L$ singlets:
\be
U_j=u_R,c_R,t_R \qquad D_j=d_R,s_R,b_R \,.
\ee
The quarks' couplings to the gluons, weak gauge bosons $W^\pm$ and $Z$, and the photon are described by the kinetic term in the Lagrangian
\be\label{eq:Lfermion}
\mathcal{L}_\text{fermion} = \sum_{j=1}^3
\bar Q_j i \slashed D_Q Q_j + \bar U_j i \slashed D_U U_j + \bar D_J i \slashed D_D D_j \vspace{-3mm}
\,,
\ee
with the covariant derivatives
\begin{eqnarray}
D_{Q,\mu} &=& \partial_\mu + i g_s T^a G^a_\mu + i g \tau^a W^a_\mu + i g' Q^Y_Q B_\mu\,,
\\
D_{U,\mu} &=& \partial_\mu + i g_s T^a G^a_\mu \hspace{1.8cm} +  i g' Q^Y_U B_\mu\,,
\\
D_{D,\mu} &=& \partial_\mu + i g_s T^a G^a_\mu \hspace{1.8cm}  +  i g' Q^Y_D B_\mu\,,\label{eq:DDmu}
\end{eqnarray}
and the hypercharges assigned as $Q^Y_Q = 1/6$, $Q^Y_U = 2/3$, $Q^Y_D = -1/3$.  $T^a (a=1,\dots,8)$ and $\tau^a (a=1,2,3)$ are the generators of $SU(3)_c$ and $SU(2)_L$, respectively, and the index $j$ runs over the three generations of quark fields. It is evident that the gauge couplings are universal for all three generations.

Flavour non-universality, on the other hand, is introduced by the quark Yukawa couplings to the Higgs field, responsible for the generation of non-zero quark masses:
\be\label{eq:Yuk}
\mathcal{L}_\text{Yuk} = \sum_{i,j=1}^3 (
- Y_{U,ij} \bar Q_{Li} \tilde{H} U_{Rj} - Y_{D,ij} \bar Q_{Li} H D_{Rj}  + h.c.)\,,
\ee
where $h.c.$ abbreviates the hermitian conjugate term. The subscripts $i,j$ are generation indices, and the dual field $\tilde H$ is given as $\tilde H = \epsilon H^* = (H^{0*},-H^-)^T$. Replacing the Higgs field $H$ by its vacuum expectation value $\langle H \rangle = (0,v)^T$, we obtain the quark mass terms
\be
\sum_{i,j=1}^3 (
- m_{U,ij} \bar u_{Li} u_{Rj} - m_{D,ij} \bar d_{Li}  d_{Rj} + h.c.)\,,
\ee
with the quark mass matrices given by $m_{U,D} = v Y_{U,D}$.

The quark mass matrices $m_{U}$ amd $m_D$ are $3\times 3$ complex matrices in flavour space with {\it a priori} arbitrary entries. They can be diagonalized by making appropriate bi-unitary field redefinitions:
\be
u_L = \hat U_L u_L^m\,,\qquad u_R = \hat U_R u_R^m \,, \qquad d_L = \hat D_L d_L^m \,, \qquad d_R = \hat D_R d_R^m\,,
\ee
with the superscript $^m$ denoting quarks in their mass eigenstate basis.

Is the SM Lagrangian invariant under these  transformations?
Unitary transformations of the right-handed quark sector are indeed unphysical, as they drop out from the rest of the Lagrangian.
However, $u_{Li}$ and $d_{Li}$ form the $SU(2)_L$ doublets $Q_{i}$ (with $i=1,2,3$). Their kinetic term gives rise to the interaction
\be\label{eq:Wint}
\frac{g}{\sqrt{2}} \bar u_{Li} \gamma_\mu W^{\mu+} d_{Li}\,.
\ee
Transformation of \eqref{eq:Wint} to the mass eigenstate basis yields
\be
\frac{g}{\sqrt{2}}  \bar u_{Li} \hat U_{L,ij}^\dagger \hat D_{L,jk} \gamma_\mu W^{\mu+}  d_{Lk}\,.
\ee
We conclude that the combination 
\be
\hat V_\text{CKM} = \hat U_L^\dagger \hat D_L
\ee
 is physical, it is called the {\it CKM matrix} \cite{Cabibbo:1963yz,Kobayashi:1973fv}. It describes the misalignment between left-handed up- and down-type quark mass eigenstates, which leads to flavour violating charged current interactions, mediated by the $W^\pm$ bosons. It is convenient to label the elements of $\hat V_\text{CKM}$ by the quark flavours involved in the respective charged current interaction:
\be
\hat V_\text{CKM} =
\begin{pmatrix}V_{ud}&V_{us}&V_{ub}\\
V_{cd}&V_{cs}&V_{cb}\\V_{td}&V_{ts}&V_{tb}\end{pmatrix}\,.
\ee
For example, the element $V_{ub}$ appears in the coupling of a bottom and an up quark to the $W$ boson.

\subsection{Standard Parametrization of the CKM Matrix}

 Let us now determine the number of physical parameters in the CKM matrix.
Being a unitary $3\times 3$ matrix, it can be parametrized by three mixing angles and six complex phases in general. However, five of these phases are unphysical, as they can be absorbed as unobservable parameters into the up-type and down-type quarks, respectively. Note that an overall phase rotation of all quarks does not affect the CKM matrix.
We are then left with  three mixing angles $\theta_{12}$, $\theta_{23}$, $\theta_{13}$ and one complex phase $\delta$ as the physical parameters of the CKM matrix. Introducing the short-hand notation $s_{ij} = \sin\theta_{ij}$ and $c_{ij} = \cos\theta_{ij}$, the standard parametrization of the CKM matrix reads \cite{Chau:1984fp}
\be
\hat V_\text{CKM} =
\begin{pmatrix}
c_{12} c_{13} & s_{12} c_{13} & s_{13} e^{-i\delta} \\
-s_{12} c_{23} - c_{12} s_{23} s_{13} e^{i\delta} & c_{12} c_{23} - s_{12} s_{23} s_{13} e^{i\delta} & s_{23} c_{13} \\
s_{12} s_{23} - c_{12} c_{23} s_{13} e^{i\delta} & - c_{12} s_{23} - s_{12} c_{23} s_{13} e^{i\delta} & c_{23} c_{13}
\end{pmatrix}\,.
\ee
Note that this parametrization is recommended by the Particle Data Group (PDG) \cite{Olive:2016xmw}.

Alternatively, the number of independent flavour parameters in the SM can also be determined from symmetry principles. Ignoring the Yukawa couplings, the SM quark sector has a global
\be
G_\text{flavour} = U(3)_Q \times U(3)_U \times U(3)_D
\ee
 flavour symmetry. The quark Yukawa couplings $Y_U$, $Y_D$ explicitly break $G_\text{flavour}$, leaving only a single $U(1)$ factor unbroken, that corresponds to the overall phase of the quark fields. This $U(1)$ symmetry is associated to baryon number conservation, which is an accidental symmetry of the SM. We can use this symmetry breaking pattern to count the number of flavour parameters.

We start from the Yukawa couplings $Y_U$ and $Y_D$. A priori, these are arbitrary complex $3\times 3$ matrices, hence they bring in nine real parameters and nine complex phases each. However, not all of these $18+18$ parameters are physical. In fact, each of the broken generators of the flavour symmetry group $G_\text{flavour}$ corresponds to an unphysical parameter in the Lagrangian which can be removed by making appropriate field redefinitions. A $3\times 3$ unitary matrix contains three real parameters and six complex phases. The three $U(3)$ factors in $G_\text{flavour}$ therefore carry nine real parameters and 18 phases. All but one of them, namely the phase corresponding to the unbroken overall $U(1)$, correspond to unphysical parameters that can be removed from  $Y_U$ and $Y_D$. We are then left with a total of nine real parameters in the quark flavour sector and one physical complex phase. The nine real parameters are the quark masses $m_u, m_d,m_c,m_s,m_t,m_b$ and the three mixing angles $\theta_{12},\theta_{13},\theta_{23}$ of the CKM matrix, and the phase is simply the CKM phase $\delta$.

Experimentally, it has been found that the CKM matrix exhibits a rather strong hierarchy, with \cite{Olive:2016xmw}
\be
s_{12} \sim 0.2\,,\qquad s_{23} \sim 0.04 \,,\qquad s_{13} \sim 4\cdot 10^{-3} \,.
\ee
The CKM matrix hence is close to the unit matrix, with hierarchical off-diagonal elements. Flavour changing transitions are therefore strongly suppressed in the SM. Similarly, also the quark masses are found to follow a hierarchical pattern, spanning five orders of magnitude in size. The lack of a more fundamental theory explaining the origin of this structure is referred to as the flavour hierarchy problem of the SM. 

\subsection{\boldmath $CP$ violation in the SM}

We have seen above that the angles $\theta_{ij}$ of the CKM matrix parametrize the amount of flavour mixing between the quarks of the generations $i$ and $j$. The amount of flavour violation in the SM is therefore quantified by the values of the CKM mixing angles. But what is the physical meaning of the presence of a complex phase $\delta$?

In order to understand this, let us consider two discrete transformations:
\bi
\item
the parity transformation
\be
P: \psi(r,t) \to \gamma^0 \psi(-r,t)
\ee
which transforms left-handed fermion fields into right-handed ones and vice versa, and
\item
the charge conjugation
\be
C: \psi \to i(\bar\psi \gamma^0\gamma^2)^T
\ee
which transforms left(right)-handed quarks into left(right)-handed antiquarks.
\ei
It is evident from equations \eqref{eq:Lfermion}--\eqref{eq:DDmu} that the weak interactions violate both $C$ and $P$, as they treat left- and right-handed quarks differently. 

But what about the combination of both transformations, $CP$? A $CP$ transformation connects left-handed quarks to right-handed antiquarks. It is easy to convince oneself that the neutral current interactions mediated by gluons, the photon and the $Z$ boson are indeed invariant under $CP$. Let us then look at the charged current interactions mediated by the $W^\pm$ bosons:
\begin{eqnarray}
\mathcal{L}_{c.c.} &=& \frac{g}{\sqrt{2}} V_{ik} \bar u_{Li}  \gamma_\mu W^{\mu+} d_{Lk} + h.c. \nn\\
 &=&  \frac{g}{\sqrt{2}}  V_{ik} \bar u_{Li} \gamma_\mu W^{\mu+} d_{Lk} + \frac{g}{\sqrt{2}} V_{ik}^* \bar d_{Lk} \gamma_\mu W^{\mu-} u_{Li} \nn\\
&\xrightarrow{CP}&  \frac{g}{\sqrt{2}}  V_{ik}  \bar d_{Lk}\gamma_\mu W^{\mu-} u_{Li} + \frac{g}{\sqrt{2}}  V_{ik}^* \bar u_{Li}   \gamma_\mu W^{\mu+} d_{Lk}\nn\\
 &=&  \frac{g}{\sqrt{2}}   V_{ik}^* \bar u_{Li} \gamma_\mu W^{\mu+} d_{Lk} + h.c.
\end{eqnarray}
We see that $CP$ conjugation replaces the CKM element $V_{ik}$ by its complex conjugate. Hence, the $CP$ symmetry is violated in the SM by the presence of a non-vanishing complex phase $\delta\ne 0$ in the CKM matrix. 

It is important to note, however, that the phase $\delta$ is not a physical parameter, as, by means of the aforementioned rephasing of quark fields, it can be shifted to different elements of the CKM matrix. A parametrization-independent and therefore physical measure of $CP$ violation is instead given by the Jarlskog invariant \cite{Jarlskog:1985ht,Jarlskog:1985cw}
\be
J_{CP} = \text{Im}(V_{us}V_{cb} V_{ub}^*V_{cs}^*)\,.
\ee
Experimentally, the Jarlskog invariant is found to be $J_{CP}\simeq 3\cdot 10^{-5}$.

\subsection{Flavour Changing Neutral Currents}\label{sec:FCNC}

We have seen above that flavour changing charged currents are present at the tree level in the SM, with the size of the interactions governed by the off-diagonal elements of the CKM matrix. Flavour changing neutral currents (FCNCs), on the other hand, are absent at the tree level in the SM. In order to see this, let us have a closer look at the $Z$ boson coupling to left-handed down-type quarks, as an illustrative example. Transforming the coupling of the quark flavour eigenstates $d_{Lj}$ into a coupling of the quark mass eigenstates $d_{Li}^m$, we find 
\begin{eqnarray}
\mathcal{L}_\text{n.c} &\ni&
g_Z(d_L)\, \bar d_{Lj}  \gamma_\mu Z^{\mu}  d_{Lj} \nn\\
&=& g_Z(d_L)\, \bar d_{Li}^m (\hat{D}_{L}^\dagger)_{ij} \gamma_\mu Z^{\mu} (\hat{D}_L)_{jk} d_{Lk}^m \nn\\
 &=& g_Z(d_L) \, \bar d_{Li}^m \gamma_\mu Z^{\mu} \delta_{ik} d_{Lk}^m \,,
\end{eqnarray}
where
\be
g_Z(d_L) = \frac{g}{\cos\theta_W}\left(-\frac{1}{2}+\frac{1}{3}\sin^2\theta_W\right)
\ee
is the $Z$ boson coupling to left-handed down-type quarks. We can see that due to the unitarity of the flavour rotation matrix $\hat D_L$, the coupling remains flavour diagonal and flavour universal. The same argument holds for all neutral gauge boson couplings in the SM. 

\begin{figure}
\center{\includegraphics[width=.25\textwidth]{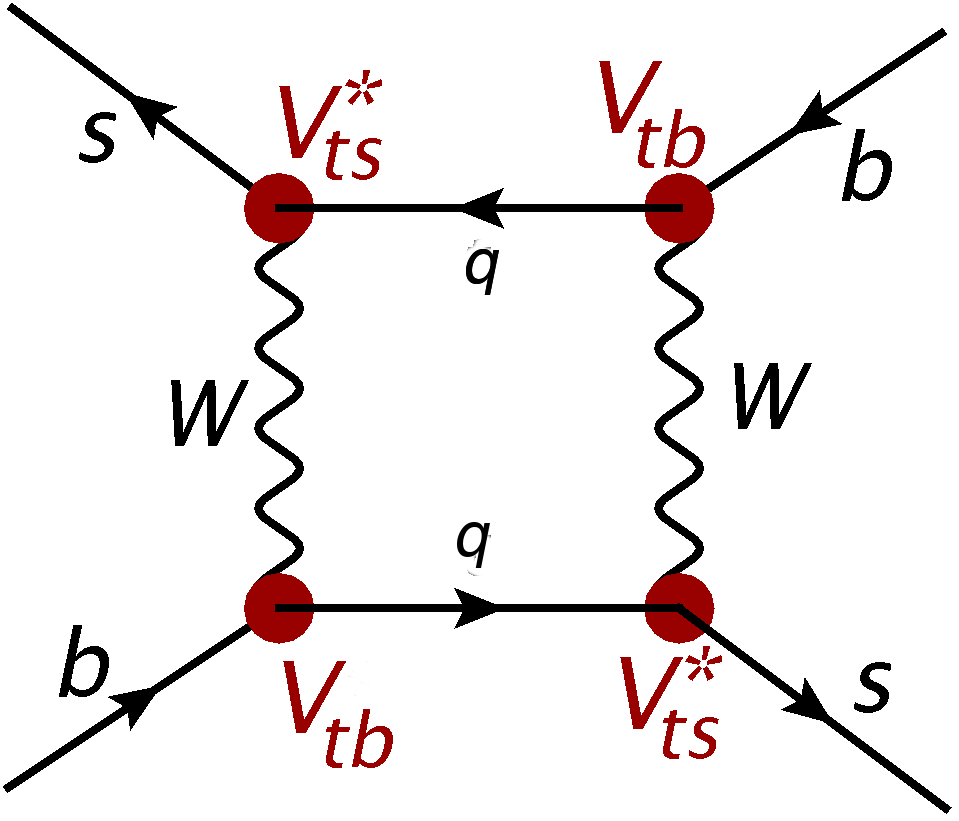}}
\caption{One loop diagram for $B_s-\bar B_s$ mixing in the SM.}\label{fig:Bs-mixing}
\end{figure}

FCNCs are however generated in the SM by loop diagrams with internal $W^\pm$ bosons. As an example, figure \ref{fig:Bs-mixing} shows the one loop diagram that generates the leading contribution to neutral $B_s$ meson mixing. The mixing amplitude generated by this contribution is schematically given by
\be
\mathcal{M} \propto \sum_{i,j=u,c,t} V_{is}^* V_{ib} V_{js}^* V_{jb} F(x_i,x_j)\,.
\ee
Here, $F(x_i,x_j)$ is the relevant one loop function, with $x_i = (m_i/M_W)^2$, and the double sum runs over the internal quark flavours. Using the CKM unitarity
\be
\sum_{i=u,c,t} V_{is}^* V_{ib} = 0\,,
\ee
setting $m_u = 0$ and neglecting contributions proportional to $V_{cs}^* V_{cb}\ll V_{ts}^* V_{tb}$,
this can be simplified to
\be
\mathcal{M} \propto (V_{ts}^* V_{tb})^2 S_0(x_t)\,,
\ee
with
\be
S_0(x_t) = F(x_t,x_t) - 2 F(0,x_t) +F(0,0)\,.
\ee
We can see that, indeed, FCNCs are generated by loop processes in the SM. However they are suppressed not only by the smallness of the off-diagonal CKM elements, but also by the so-called {\it GIM mechanism}~\cite{Glashow:1970gm}: All contributions that are independent of the masses of the quarks running in the loop are cancelled by the unitarity of the CKM matrix, and only the differences of mass-dependent terms survive. While above we have seen the GIM mechanism at work for one loop contributions, it in fact holds to all orders.

\subsection{The Unitarity Triangle}

The hierarchical structure of the CKM matrix can be used to derive an alternative parametrization, which turns out to be very useful for estimating the size of flavour violating transitions. In the Wolfenstein parametrization \cite{Wolfenstein:1983yz}
\be
\hat V_\text{CKM} =
\begin{pmatrix}
1-\frac{\lambda^2}{2} & \lambda & A \lambda^3 (\rho-i\eta) \\
-\lambda & 1-\frac{\lambda^2}{2} & A \lambda^2 \\
A \lambda^3 (1-\rho-i\eta) & - A\lambda^2 & 1
\end{pmatrix}+\mathcal{O}(\lambda^4)\,,
\ee
 $\lambda=|V_{us}| \sim 0.2$ is the only small parameter, while $A$, $\rho$, and $\eta$ are $\mathcal{O}(1)$. It is therefore convenient to estimate the size of flavour violating decays by making an expansion in powers of $\lambda$. The accuracy of this expansion can be improved by changing the parameters of the Wolfenstein parametrization to \cite{Buras:1994ec,Schmidtler:1991tv}
\be
\lambda\,,\qquad A\,,\qquad \bar\rho = \left(1-\frac{\lambda^2}{2}\right)\rho\,,\qquad \bar\eta = \left(1-\frac{\lambda^2}{2}\right)\eta\,.
\ee

\begin{figure}
\center{\includegraphics[width=.5\textwidth]{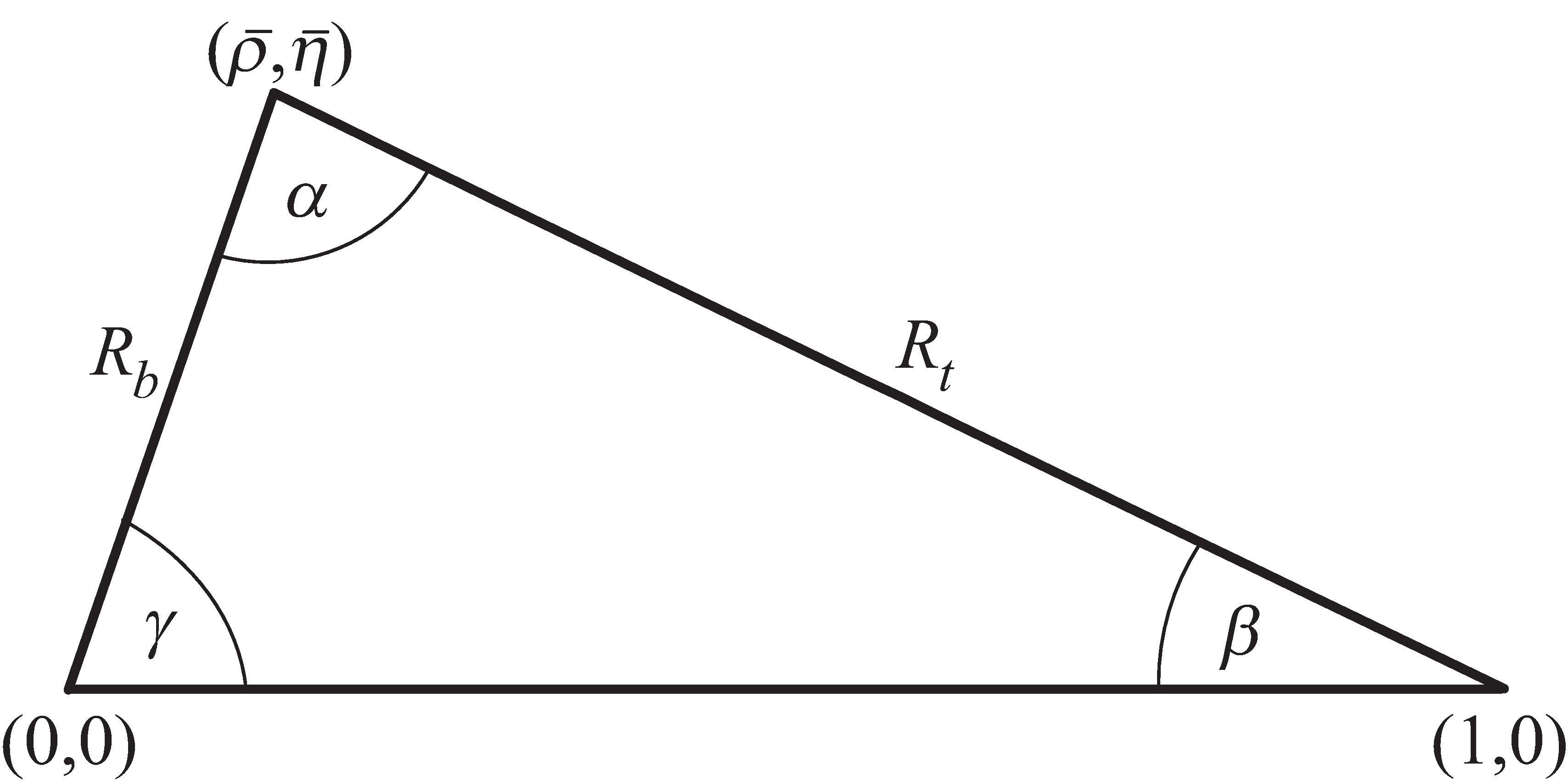}}
\caption{Unitarity triangle.}\label{fig:UT}
\end{figure}

As discussed before, the CKM matrix is a unitary matrix, and not all of its elements are independent parameters. Various relations hold among them, which can be tested experimentally.
One of the most popular ones,
\be
V_{ud}^{}V_{ub}^*+V_{cd}^{}V_{cb}^*+V_{td}^{}V_{tb}^*=0\,,
\ee
can be displayed as a triangle in the complex plane, the so-called {\it unitarity triangle} (UT) \cite{Jarlskog:1988ii}. With the base of the UT normalized to unity,
the apex is simply given by $(\bar\rho,\bar\eta)$. The sides $R_b$ and $R_t$, as shown in figure \ref{fig:UT}, are given by
\begin{eqnarray}
R_b&=&\left|\frac{V_{ud}^{}V_{ub}^*}{V_{cd}^{}V_{cb}^*}\right|
= \left(1-\frac{\lambda^2}{2}\right) \frac{1}{\lambda}\frac{|V_{ub}|}{|V_{cb}|}\,,\\
R_t&=&\left|\frac{V_{td}^{}V_{tb}^*}{V_{cd}^{}V_{cb}^*}\right|=\frac{1}{\lambda}\frac{|V_{td}|}{|V_{cb}|}\,.
\end{eqnarray}

For the UT angles, two notations are commonly used in the literature. They are related to each other as follows:
\be
\alpha \equiv \phi_2\,,\qquad \beta \equiv \phi_1\,, \qquad \gamma \equiv \phi_3\,.
\ee

The UT can be determined experimentally from various measurements of flavour violating decays of $K$ and $B$ mesons. A special role in this determination is played by the length of the side $R_b$ and the angle $\gamma$: Being sensitive to the absolute values and $CP$-violating phases of the elements $V_{ub}$ and $V_{cb}$, they can be determined from $B$ decays governed by tree level charged current interactions. 
It is therefore a good approximation to assume that NP contributions to these measurements are negligible. The measurement of $|V_{ub}|$, $|V_{cb}|$ and $\gamma = \arg(-V_{ud}^{}V_{ub}^*/V_{cd}^{}V_{cb}^*)$ then leaves us with the {\it reference unitarity triangle} \cite{Grossman:1997dd}, which determines the CKM matrix independently of potential NP contributions to rare flavour violating decays.

The length of the side $R_t$ and the angle $\beta$, on the other hand, depend on CKM elements involving the top quark. Hence, they can only be measured in loop-induced flavour changing neutral current (FCNC) processes. Due to their strong suppression in the SM, these observables are sensitive to NP contributions. A model-independent determination of the CKM matrix using these quantities is therefore not possible. NP contributions to the loop induced processes used in the determination of the UT

The strategy to hunt for NP contributions to flavour violating observables is then as follows. First, the CKM matrix and the UT have to be determined from tree level charged current decays as accurately as possible. As this determination is independent of potential NP contributions, the result can be used as input for precise SM predictions of rare, loop-induced FCNC processes. These predictions are then to be compared with the data, which -- in case of a discrepancy -- would yield an unambiguous sign of a NP contribution to the decay in question. Clearly, in order to be able to claim a NP discovery in flavour violating observables, a solid understanding of the SM contribution and its uncertainties is mandatory.

\subsection{The effective Hamiltonian}

An important theoretical complication arises in the study of quark flavour violating decays. Due to the confinement of QCD at low energies, quarks do not appear as free particles in nature, but are bound in hadrons. Therefore, not only the weak interactions leading to flavour violation, as discussed above, have to be well understood, but also the strong dynamics describing the bound states of QCD. The latter interactions, taking place at the typical hadronic energy scales of a few hundred MeV to a few GeV, are non-perturbative and hence, with our current methods, cannot be calculated analytically. 

A convenient theoretical tool to handle these various contributions from processes ar different energy scales is provided by the operator product expansion \cite{Wilson:1969zs}. In this framework, effective flavour violating operators are obtained from integrating out the heavy electroweak (EW) gauge bosons $W^\pm$ and $Z$ and the top quark at the EW scale, and then connecting these operators with the low energy QCD interactions responsible for hadronic interactions. The latter are comprised in matrix elements of the effective operators, involving the initial and final state mesons of the decay in question. These matrix elements, being governed by non-perturbative interactions, cannot be calculated analytically, but have to evaluated using non-perturbative methods like lattice QCD or QCD sum rules (see e.\,g.\ \cite{Lellouch:2011qw} and \cite{Radyushkin:1998du} for pedagogical introductions)
, unless it is possible to extract them from the data.  

To summarize, in order to arrive at a theoretical description of flavour violating meson decays, the following five steps have  to be taken:
\begin{enumerate}
\item
Calculation of the weak interaction process governing the underlying flavour violating quark decay.
\item
Construction of the low energy effective Hamiltonian by integrating out the heavy degrees of freedom (i.\,e.\ $W^\pm$, $Z$, $t$).
\item
Renormalization group running from the scale $\mu\sim M_W$ of weak interactions to the hadronic scale $\mu\sim\text{GeV}$.
\item
Collection of non-perturbative effects in QCD matrix elements involving initial and final state mesons.
\item
Evaluation of matrix elements using non-perturbative methods (lattice QCD, QCD sum rules etc.) or extraction from data.
\end{enumerate}

To better understand how this is done in practice, let us have a look at two simple examples. First we consider the semileptonic charged current decay $B\to\pi\ell\nu$ from which the CKM element $|V_{ub}|$ can be obtained. Then we sketch the SM prediction for neutral $B_s$ meson mixing.

\subsubsection*{Semileptonic charged currents: $B\to\pi\ell\nu$}

\begin{figure}[h]
\center{\includegraphics[width=.25\textwidth]{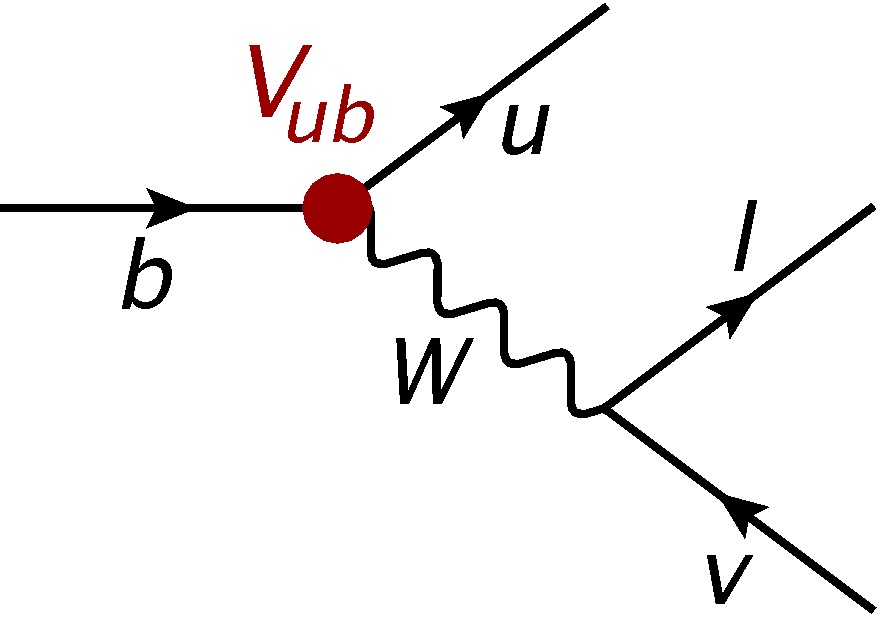}}
\caption{Tree level diagram mediating the $b\to u\ell\nu$ decay.}\label{fig:bulnu}
\end{figure}

The element $|V_{ub}|$ can be determined from the semileptonic charged current transition $b\to u \ell\nu$, occurring at the tree level in the SM. Evaluating the relevant Feynman diagram shown in figure \ref{fig:bulnu}, we find
\be
\mathcal{M} = \frac{g}{\sqrt{2}} V_{ub} (\bar u\gamma^\mu P_L b) \frac{g_{\mu\nu}}{p^2-M_W^2} \frac{g}{\sqrt{2}} (\bar\ell\gamma^\nu P_L \nu)\,.
\ee
Neglecting the momentum transfer $p^2 \ll M_W^2$ in the $W$ propagator, and using the short hand notation $2 (\bar f \gamma^\mu P_L f') = (\bar f f')_{V-A}$, $\mathcal{M}$ simplifies to
\be
\mathcal{M} = -\frac{g^2}{2M_W^2} V_{ub} (\bar ub)_{V-A} (\bar\ell\nu)_{V-A}\,.
\ee
Introducing the well-known Fermi constant
\be
\frac{G_F}{\sqrt{2}} = \frac{g^2}{8M_W^2}\,,
\ee
we obtain the tree level effective Hamiltonian
\be
\mathcal{H}_\text{eff} = \frac{4G_F}{\sqrt{2}} V_{ub} (\bar ub)_{V-A} (\bar\ell\nu)_{V-A} + h.c. \,.
\ee

In order to obtain an accurate expression for the $B\to\pi\ell\nu$ decay, the known QCD corrections have to be taken into account, including the renormalization group running from the weak scale, where the $W$ boson is integrated out, to the $B$ meson scale. The non-perturbative QCD effects describing the $B\to\pi$ transition generated by the effective Hamiltonian $\mathcal{H}_\text{eff}$ are collected in the matrix element
\be
\langle\pi|(\bar bu)_{V-A}|B\rangle  \,,
\ee
which has been calculated by lattice QCD. 

From the measurement of the $B^0\to\pi^-\ell^+\nu$ branching ratio, one can then extract the value \cite{Lattice:2015tia} (see also \cite{Dalgic:2006dt,Flynn:2015mha})
\be\label{eq:Vub}
|V_{ub}| = (3.72 \pm 0.16) \cdot 10^{-3}\,.
\ee
Note again that as the $B^0\to\pi^-\ell^+\nu$ is generated at the tree level in the SM, NP contributions are expected to be negligible. Therefore the determination of $|V_{ub}|$ in \eqref{eq:Vub} holds model-independently. In a similar way, also $|V_{us}|$, $|V_{cb}|$ and the UT angle $\gamma$ can be determined from tree level decays, independently of NP in flavour violating decays.

\subsubsection*{$B_s-\bar B_s$ mixing}

As discussed in section \ref{sec:FCNC}, neutral $B_s$ meson mixing in the SM is driven by the one loop box diagram in figure \ref{fig:Bs-mixing}. We have seen that due to the GIM mechanism and the hierarchical structure of the CKM matrix, only the mass-dependent part of the top quark contribution  is relevant. Including a factor $\eta_B =0.55\pm0.01$ \cite{Buras:1990fn,Urban:1997gw} that comprises perturbative QCD corrections and the renormalization group evolution down to the $B$ meson scale, the effective Hamiltonian can be written as
\be
\mathcal{H}_\text{eff} = \frac{G_F^2}{16\pi^2}M_W^2 \eta_B (V_{tb}^* V_{ts})^2 S_0(x_t) (\bar bs)_{V-A} (\bar bs)_{V-A} +h.c.\,,
\ee
with the one loop function $S_0(x_t)$ given by
\be
S_0(x_t) = \frac{4 x_t-11x_t^2+x_t^3}{4(1-x_t)^2}-\frac{3x_t^3 \ln x_t}{2(1-x_t)^3}\,.
\ee
Sandwiching $\mathcal{H}_\text{eff} $ between the initial and final state mesons, we obtain the mixing matrix element
\be
M_{12} = \frac{1}{2 m_{B_s}} \left\langle \bar B_s \left| \mathcal{H}_\text{eff} \right| B_s \right\rangle^*\,.
\ee
Again, the hadronic matrix element $\left\langle \bar B_s \left| (\bar bs)_{V-A} (\bar bs)_{V-A} \right| B_s \right\rangle$ comprises the low-energy QCD dynamics of the mesonic process. It has been calculated by lattice QCD with an impressive precision. 

Using the CKM matrix determined from tree level decays, the SM predictions for $B_s-\bar B_s$ mixing observables, like the mass difference $\Delta M_s$ and the $CP$-violating phase $\phi_s$ can then be compared with the data. This comparison yields a good agreement of the measured values with their SM predictions, leaving only little room for NP contributions. In lecture \ref{sec:NP} we will have a closer look at the constraints on NP models from flavour violating observables.

\section{\boldmath Phenomenology of $K$ and $B$ Meson Decays}\label{sec:pheno}

\subsection{Theory of Neutral Kaon Mixing}

In the first lecture, we have briefly sketched the theoretical description of $B_s - \bar B_s$ mixing in the SM, paying particular attention to specific details like the GIM mechanism and the construction of the effective Hamiltonian. In this section, we aim at a more thorough derivation of neutral meson mixing. While we focus on the case of kaon mixing in what follows, a generalization to neutral $B$ and $D$ mesons is straightforward. 

Two neutral pseudoscalar $K$ mesons exist, the $|K^0\rangle = |d\bar s\rangle$ and the $|\bar K^0\rangle = |s\bar d\rangle$. They are each other's antiparticles and transform under $CP$ as
\be
CP | K^0\rangle = -| \bar K^0\rangle\,, \qquad
CP | \bar K^0\rangle = -| K^0\rangle \,.
\ee

As we can deduce from section \ref{sec:FCNC}, $K^0$ and $\bar K^0$ can mix by one loop box diagrams in the SM. The time evolution of the $K^0-\bar K^0$ system is therefore described  by the two-component Schr\"odinger equation
\be
i \frac{d\psi(t)}{dt} = \hat H \psi(t)\,,\qquad \psi(t) = \begin{pmatrix} K^0(t)\\ \bar K^0(t) \end{pmatrix}
\ee
with the $2\times 2$ Hamiltonian
\be
\hat H = \hat M -i\frac{\hat\Gamma}{2}=\begin{pmatrix}
M_{11} & M_{12} \\
M_{21} & M_{22}\end{pmatrix}
-\frac{i}{2}\begin{pmatrix}
\Gamma_{11} & \Gamma_{12} \\
\Gamma_{21} & \Gamma_{22} 
\end{pmatrix}\,.
\ee
Here, $\hat M$ is the dispersive part of the Hamiltonian, and $\hat\Gamma$ is the absorptive part. Since both $\hat M$ and $\hat\Gamma$ are hermitian, we have
\be
M_{21} = M_{12}^*\,,\qquad\Gamma_{21} = \Gamma_{12}^*\,.
\ee
In addition, $CPT$ invariance implies
\be
M_{11}=M_{22}\equiv M\,,\qquad \Gamma_{11}=\Gamma_{22}\equiv \Gamma\,,
\ee
so that the Hamiltonian simplifies to
\be
\hat H = \begin{pmatrix}
M-\frac{i}{2}\Gamma & M_{12}-\frac{i}{2}\Gamma_{12} \\
M^*_{12}-\frac{i}{2}\Gamma_{12}^* & M-\frac{i}{2}\Gamma 
\end{pmatrix}
\ee

Diagonalizing $\hat H$, we obtain the mass eigenstates
\be\label{eq:K-mass}
K_{L,S} = \frac{(1+\bar\varepsilon)K^0\pm (1-\bar\varepsilon)\bar K^0}{\sqrt{2(1+|\bar\varepsilon|^2)}}\,,
\ee
where $L$ and $S$ stand for `long' and `short', respectively, and refer to the different lifetimes of the two states. The parameter  $\bar\varepsilon$ is defined through
\be
\frac{1-\bar\varepsilon}{1+\varepsilon}=\sqrt{\frac{M^*_{12}-\frac{i}{2}\Gamma_{12}^*}{M_{12}-\frac{i}{2}\Gamma_{12}}}=
\frac{\Delta M -\frac{i}{2}\Delta\Gamma}{2M_{12}-i\Gamma_{12}}\,.
\ee

Experimentally $\bar\varepsilon$ is found to be of the order $\mathcal{O}(10^{-3})$. Therefore the mass difference $\Delta M $ and the width difference $\Delta \Gamma$ are well approximated by the simple expressions
\be
\Delta M = M_L-M_S = 2\,\text{Re} M_{12}\,,\qquad \Delta \Gamma = \Gamma_L-\Gamma_S = 2\,\text{Re}\Gamma_{12}\,.
\ee

\subsection{$CP$ Violation in the Neutral Kaon Sector}

As we have seen before, $K^0$ and $\bar K^0$ transform into each other under $CP$ conjugation.  The CP eigenstates are therefore given by
\be\label{eq:K-CP}
K_1 = \frac{1}{\sqrt{2}}(K^0-\bar K^0)\,, \qquad 
K_2 = \frac{1}{\sqrt{2}}(K^0+\bar K^0)\,.
\ee
$K_1$ is even under $CP$ conjugation, while $K_2$ is $CP$-odd. Comparing the $CP$ eigenstates in \eqref{eq:K-CP} to the mass eigenstates in \eqref{eq:K-mass}, it becomes clear that the mass eigenstates $K_{L,S}$ are not pure $CP$ eigenstates. Instead they contain a small admixture of the state with opposite $CP$ parity:
\be
K_S=\frac{K_1+\bar\varepsilon K_2}{\sqrt{1+|\bar\varepsilon|^2}}\,,\qquad
K_L=\frac{K_2+\bar\varepsilon K_1}{\sqrt{1+|\bar\varepsilon|^2}}\,.
\ee
This mixture of states with opposite $CP$ parities implies that the $CP$ symmetry is violated in neutral kaon mixing.

The two mass eigenstates, $K_L$ and $K_S$ are found experimentally to have very different lifetimes~\cite{Olive:2016xmw}:
\be
\tau(K_S) \sim 90\,\text{ps}\,, \qquad \tau(K_L) \sim 5\cdot 10^3\,\text{ps}\,.
\ee
The explanation can be found in the $CP$ properties of the two states: $K_S$ is basically the $CP$-even state $K_1$, with a small $CP$-odd admixture. $K_L$, on the other hand, is approximately the $CP$-odd state $K_2$, with a small $CP$-even admixture. Now if $CP$ is conserved in the decay of neutral kaons, then the $CP$-even state $K_1$ will decay into two pions, forming a $CP$-even final state. $K_2$, however, has to decay into a $CP$-odd final state, which contains three pions. As the three pion final state is phase space suppressed with respect to the two pion final state, the $K_1$ decay rate is much faster, so that its lifetime is much shorter than the one of $K_2$. The observed lifetime difference suggests that indeed $CP$ is, at least approximately, conserved and $\bar\varepsilon$ is small.

In 1964, the decay $K_L\to\pi^+\pi^-$ has been observed \cite{Christenson:1964fg}, yielding the first experimental confirmation that $CP$ symmetry is violated. In 1980, the Nobel Prize has been awarded to Cronin and Fitch for this discovery.

The mere discovery of the $K_L\to\pi^+\pi^-$ decay however does not tell us where the observed $CP$ violation originates from. $CP$ can either be violated in the neutral kaon mixing, if the mass eigenstates $K_{L,S}$ are not CP eigenstates. Or $CP$ can be violated in the decay -- in that case the $CP$-odd state $K_2$ can decay into a $CP$-even two pion final state. Last but not least, in general, $CP$ can also be violated in the interference of mixing and decay amplitudes. 

How can we distinguish whether $CP$ is violated in the mixing (also called indirect $CP$ violation) or in the decay process (direct $CP$ violation)? The key idea is that the amount of direct $CP$ violation depends on the decay channel, but indirect $CP$ violation does not.

Hence, to disentangle the two types of $CP$ violation, we study the following set of $K\to\pi\pi$ decay modes:
\begin{align}
K_L\to\pi^0\pi^0\,, \qquad K_L\to\pi^+\pi^-\,,\\
K_S\to\pi^0\pi^0\,, \qquad  K_S\to\pi^+\pi^-\,.
\end{align}
As the charged and neutral pions form an isospin triplet, the two pion final state can have either isospin $I=0$ or $I=2$. The decay amplitudes into charged and neutral pions can therefore be writen as
\begin{eqnarray}
A(K^0\to\pi^+\pi^-) &=& \sqrt{\frac{2}{3}} A_0 e^{i\delta_0} + \sqrt{\frac{1}{3}} A_2 e^{i\delta_2} \,,\\
A(K^0\to\pi^0\pi^0) &=& \sqrt{\frac{2}{3}} A_0 e^{i\delta_0} - 2 \sqrt{\frac{1}{3}} A_2 e^{i\delta_2}\,.
\end{eqnarray}
Here, $A_0$ and $A_2$ parametrize the decay amplitudes in the $I=0$ and $I=2$ final states, respectively. The `strong phases' $\delta_0$ and $\delta_2$ do not change sign under $CP$ conjugation.

Defining then the ratios
\be
\eta_{00} = \frac{A(K_L\to\pi^0\pi^0)}{A(K_S\to\pi^0\pi^0)} \,,\qquad
\eta_{+-} = \frac{A(K_L\to\pi^+\pi^-)}{A(K_S\to\pi^+\pi^-)}\,,
\ee
it is possible to disentangle $CP$ violation in neutral kaon mixing, parametrized by $\varepsilon$, from direct $CP$ violation in the $K\to\pi\pi$ decays, parametrized by $\varepsilon'$. One can show that
\be
\varepsilon \simeq \frac{1}{3}\left(\eta_{00}+2\eta_{+-}\right)\,, \qquad
\text{Re}\left(\frac{\varepsilon'}{\varepsilon}\right) \simeq \frac{1}{6}\left(1-\left|\frac{\eta_{+-}}{\eta_{00}}\right|^2\right)\,.
\ee

\subsection{Status of $\varepsilon$ and $\varepsilon'$}

Both $|\varepsilon|$ and Re$(\varepsilon'/\varepsilon)$ have been measured with high precision, with the results \cite{Olive:2016xmw}
\begin{eqnarray}
|\varepsilon| &=& (2.228\pm 0.011)\cdot 10^{-3} \,,\\
\text{Re}(\varepsilon'/\varepsilon) &=& (16.6\pm 2.3)\cdot 10^{-4}\,.\label{eq:epe-exp}
\end{eqnarray}

In the SM, $CP$ violation in the kaon sector is strongly suppressed. As the presence of three quark generations is needed for $CP$ violation, both $\varepsilon$ and $\varepsilon'$ are generated by top quark contributions. The effect is therefore proportional to the combination of CKM elements
\be
\text{Im}(V_{ts}^* V_{td}) \simeq \mathcal{O}(10^{-4})\,.
\ee
$CP$ violation in the kaon sector is thus strongly suppressed in the SM. In the presence of NP, however, this strong CKM suppression can be absent, depending on the flavour structure of the model. $CP$-violating observables in the kaon sector therefore have an outstanding sensitivity to NP contributions. 

For the parameter $\varepsilon$ a simple yet precise formula can be derived:
\be
\varepsilon = \frac{\kappa_\varepsilon e^{i\varphi_\varepsilon}}{\sqrt{2}\Delta M_K} \text{Im} M_{12}\,.
\ee
Here, the mass splitting $\Delta M_K=(0.5292\pm0.0009)\cdot 10^{-2}\,\text{ps}^{-1}$ and the phase $\varphi_\varepsilon=43.51^\circ$ have been measured precisely \cite{Olive:2016xmw}. The parameter $ \kappa_\varepsilon$ comprises corrections from long-distance dynamics, and has been estimated to be
$\kappa_{\varepsilon} = 0.94\pm 0.02$ \cite{Buras:2008nn,Buras:2010pza}. 

The off-diagonal element $M_{12}$ of the mixing amplitude is, as discussed above, generated by box diagrams in the SM. While its real part receives sizeable long-distance contributions, the $CP$-violating imaginary part is driven by short distance dynamics and therefore under good theoretical control. Including the known higher oder perturbative corrections and the non-perturbative parameter $\hat B_K$ obtained from lattice QCD calculations, one finds \cite{Brod:2011ty}
\be
|\varepsilon|_\text{SM} = (1.90\pm0.26)\cdot 10^{-3}\,,
\ee
which is a bit lower albeit still consistent with the data. 

Due to its strong suppression in the SM, $\varepsilon$ is very sensitive to potential NP contributions. The good agreement of the measured value with its SM prediction therefore results in strong constraints on the NP entering $K^0-\bar K^0$ mixing. We will return to this topic in more detail in the third lecture.

The ratio $\varepsilon'/\varepsilon$ has recently received a lot of attention. While its measured value in \eqref{eq:epe-exp} has been available since the late 1990s, until recently no reliable SM prediction was available. The situation changed when the first lattice QCD calculations of the relevant $K\to\pi\pi$ hadronic matrix elements became available. The result reads \cite{Bai:2015nea}
\be\label{eq:B6B8}
B_6^{(1/2)} = 0.57\pm 0.19\,, \qquad B_8^{(3/2)} = 0.76\pm 0.05\,.
\ee
While this result, in particular the one for $B_6^{(1/2)}$, still carries sizeable uncertainties, it is interesting to note its consistency with the bound
\be
B_6^{(1/2)} < B_8^{(3/2)} < 1\,,
\ee
that has recently been derived using large $N_c$ counting and the dual QCD approach \cite{Buras:2015xba,Buras:2016fys}.

The result for the hadronic matrix elements in \eqref{eq:B6B8} can then be plugged into the simple phenomenological expression for $\varepsilon'/\varepsilon$ in the SM \cite{Buras:2003zz,Buras:2015yba}:
\be\label{eq:epe-phen}
\text{Re}(\varepsilon'/\varepsilon)_\text{SM} =\frac{\text{Im}(V_{ts}^* V_{td})}{1.4\cdot 10^{-4}}\cdot 10^{-4} \cdot\Big(
{-3.6+21.4 B_6^{(1/2)}}+
{1.2-10.4 B_8^{(3/2)}}
\Big)\,,
\ee
which has been derived using the calculation of perturbative QCD contributions at next-to-leading order (NLO).  

The first two terms in the brackets of \eqref{eq:epe-phen} stem from the $I=0$ amplitude which is dominantly generated by QCD penguin contributions. The last two terms, on the other hand, originate in the $I=2$ amplitude, caused mainly by EW penguin contributions. The numerical values of these contributions, together with the result for the hadronic matrix elements, leads to a large cancellation between the $I=0$ and $I=2$ contributions to $\text{Re}(\varepsilon'/\varepsilon)_\text{SM}$. Due to this cancellation, a precise knowledge of the hadronic matrix elements $B_6^{(1/2)}$ and $B_8^{(3/2)}$ is of utmost importance for an accurate prediction of $\text{Re}(\varepsilon'/\varepsilon)$ in the SM. Using the result in \eqref{eq:B6B8}, one finds \cite{Buras:2015yba}
\be\label{eq:epe-SM}
\text{Re}(\varepsilon'/\varepsilon)_\text{SM} = (1.4\pm 4.6)\cdot 10^{-4}\,.
\ee 
This prediction is significantly lower than the measured value in \eqref{eq:epe-exp}, revealing a $2.9\sigma$ tension. A consistent result has been obtained in \cite{Kitahara:2016nld}.
It is interesting to note that the central value in \eqref{eq:epe-SM} is much lower than the long-standing result in \cite{Pallante:2001he}, although consistent due to the large uncertainties in the latter analysis.  We are looking forward to future
improved lattice QCD calculations by several groups which will clarify the present
situation and hopefully strengthen the indicated hint for NP.

Having at hand only a single lattice result for the hadronic matrix elements in question, it would clearly be premature to claim the presence of NP in $\varepsilon'/\varepsilon$. The observed tension, however, is intriguing. Due to its strong suppression in the SM, the ratio $\varepsilon'/\varepsilon$ is extremely sensitive to NP contributions. It is therefore conceivable that NP would first be observed in this observable, even if other flavour data measured so far show little or no discrepancy with their SM predictions. 

Following the recent progress on the theoretical understanding of $\varepsilon'/\varepsilon$, this observable has been revisited in the context of various NP models \cite{Blanke:2015wba,Buras:2015yca,Buras:2015jaq,Buras:2015kwd,Buras:2016dxz,Tanimoto:2016yfy,Kitahara:2016otd,
Bobeth:2016llm}. It turns out that several extensions of the SM can significantly enhance $\varepsilon'/\varepsilon$ and thereby reconcile the theory prediction with the data. In addition, many NP scenarios predict simultaneous large deviations from the SM prediction of the rare decay $K_L\to\pi^0\nu\bar\nu$, with the sign of the latter effect depending on the structure of the model.

\subsection{\boldmath Rare $K$ Decays}

Rare and $CP$-violating kaon decays, like the aforementioned decay $K_L\to\pi^0\nu\bar\nu$, offer a unique opportunity to look for NP. These decays, mediated by $s\to d$ FCNC transitions at the quark level, are strongly suppressed in the SM by the hierarchical structure of the CKM matrix and the GIM mechanism.  Consequently, large NP effects are possible even if the NP mass scale is much beyond the TeV scale. 
Of particular interest are the decay modes $K^+\to\pi^+\nu\bar\nu$ and $K_L\to\pi^0\nu\bar\nu$, as they are not only strongly suppressed in the SM but also theoretically extremely clean. Therefore, an outstanding NP sensitivity is provided by these decays. In fact, it has been shown within a simplified model analysis that a flavour violating $Z'$ gauge boson can lead to large effects in the $K\to\pi\nu\bar\nu$ decays even if its mass is in the $10^3$\,TeV range \cite{Buras:2014zga}. 

\begin{figure}[h]
\center{\includegraphics[width=.27\textwidth]{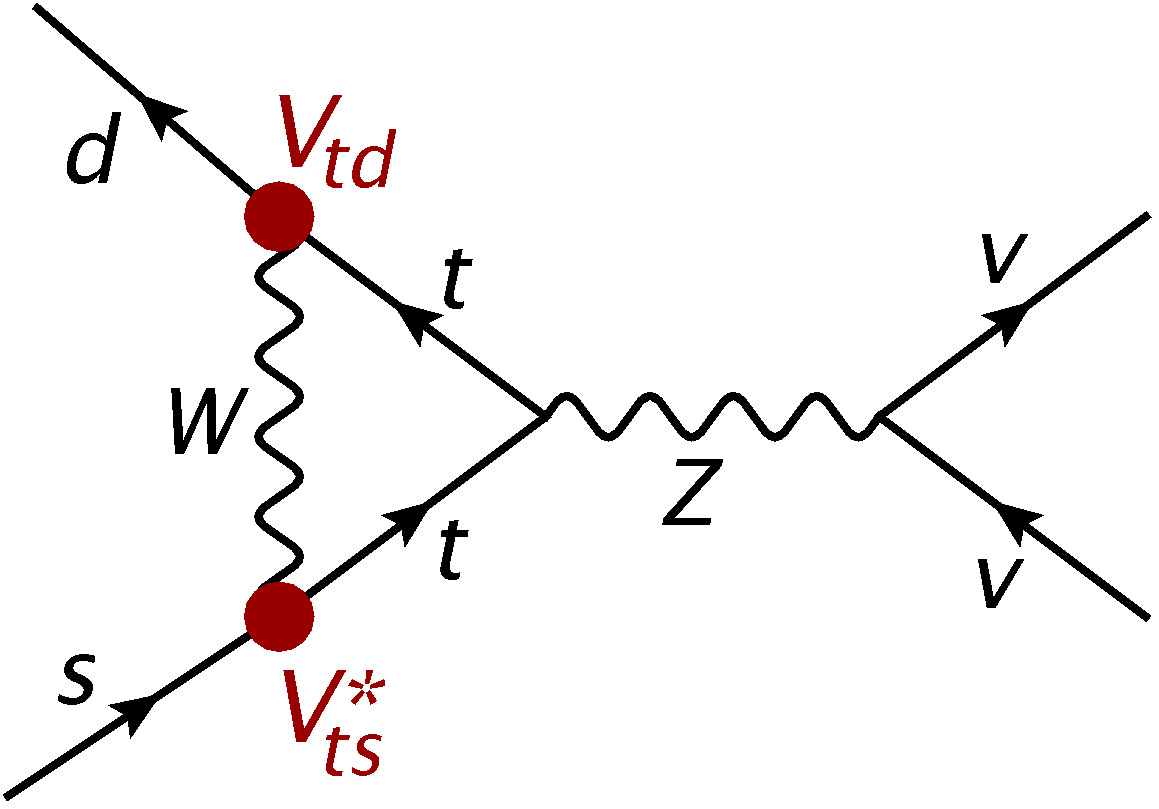}}
\caption{$Z$-penguin diagram contributing to $K\to\pi\nu\bar\nu$ in the SM.}\label{fig:kpnn-SM}
\end{figure}

In the SM, the $K\to\pi\nu\bar\nu$ decays are governed by $Z$-penguin and box diagrams like the ones shown in figure \ref{fig:kpnn-SM}. The effective Hamiltonian reads \cite{Buchalla:1995vs}
\be\label{eq:Heff-kpnn}
\mathcal{H}_\text{eff} = \frac{G_F}{\sqrt{2}}\frac{\alpha}{2\pi\sin^2\theta_W} \sum_{\ell=e,\mu,\tau} \left(V_{cs}^* V_{cd} X^\ell_\text{NNL}(x_c) + V_{ts}^* V_{td} X(x_t) \right) (\bar s d)_{V-A} (\bar \nu_\ell \nu_\ell)_{V-A}\,.
\ee
The first term in the brackets corresponds to the charm quark contribution which is known to NNLO in QCD \cite{Buras:2005gr,Buras:2006gb} and NLO in the EW theory \cite{Brod:2008ss}. It is relevant only for the $CP$-conserving decay $K^+\to\pi^+\nu\bar\nu$. The second term stems from the top quark contribution which affects both the $CP$-conserving mode $K^+\to\pi^+\nu\bar\nu$ and the $CP$-violating mode $K_L\to\pi^0\nu\bar\nu$.

The relevant $\langle \pi | (\bar s d)_{V-A} | K \rangle$ matrix elements can be extracted from the data on $K^+\to\pi^0 e^+\nu$ with high precision, making use of isospin symmetry. The main uncertainties is the SM prediction for the $K\to\pi\nu\bar\nu$ branching ratios therefore stem from the determination of the relevant CKM elements. Of particular importance is the value of $|V_{cb}|$ and, for $K_L\to\pi^0\nu\bar\nu$, the UT angle $\gamma$. The current SM prediction for the two branching ratios is \cite{Buras:2015qea}
\begin{eqnarray}
\mathcal{B}(K^+\to\pi^+\nu\bar\nu) &=& (8.4 \pm 1.0) \cdot 10^{-11}\,,\\
\mathcal{B}(K_L\to\pi^0\nu\bar\nu) &=& (3.4 \pm 0.6) \cdot 10^{-11}\,.
\end{eqnarray}
A big experimental effort to measure these decays is currently underway at the NA62 experiment \cite{Rinella:2014wfa} at CERN and the KOTO experiment \cite{Yamanaka:2012yma} at J-PARC in Japan.

NP contributions to $\mathcal{H}_\text{eff}$ in \eqref{eq:Heff-kpnn} can be parametrized model-independently by replacing the SM top-loop funcion $X(x_t)$ by a general complex function \cite{Buras:2004ub}
\be
X\equiv |X| e^{i\theta_X}\,.
\ee
NP in the $K\to\pi\nu\bar\nu$ system can therefore be described by two independent parameters $|X|$ and $\theta_X$. Measuring both $\mathcal{B}(K^+\to\pi^+\nu\bar\nu)$ and $\mathcal{B}(K_L\to\pi^0\nu\bar\nu)$ determines both parameters, and observing
\be
|X| \ne X(x_t) \qquad \text{and/or} \qquad \theta_X \ne 0
\ee
would be an unambiguous sign of NP. 

\begin{figure}
\center{\includegraphics[width=.4\textwidth]{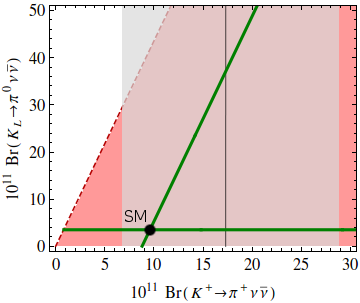}}
\caption{Model-distinguishing correlation between $\mathcal{B}(K^+\to\pi^+\nu\bar\nu)$ and $\mathcal{B}(K_L\to\pi^0\nu\bar\nu)$ \cite{Blanke:2009pq}.}\label{fig:kpnn-branches}
\end{figure}

Determining both $|X|$ and $\theta_X$ not only provides a clean test of the SM, but in case of a non-vanishing NP contribution also allows to draw conclusions about the structure of NP contributions to neutral kaon mixing \cite{Blanke:2009pq}. The reason is quite simple to understand. If the effective flavour changing $s\to d$ transition is, as in the SM, purely left-handed, then the same NP structure is responsible for $K^0-\bar K^0$ mixing and for the $K\to\pi\nu\bar\nu$ decays. In particular, the same $CP$-violating phase $\theta_K$ enters, only mutliplied by a factor of two for $K^0-\bar K^0$ mixing. The constraint on NP from $\Delta M_K$ is much weaker than the one from $\varepsilon$, so that any NP contribution to $K^0-\bar K^0$ mixing must be predominantly real: $2\theta_K \simeq 0,\pi$. The phase $\theta_K$ measured in 
the $K\to\pi\nu\bar\nu$ decays is then restricted to the values
\be
\theta_K \simeq 0,\frac{\pi}{2}, \pi, \frac{3\pi}{2}\,.
\ee
In the plane showing the branching ratios $\mathcal{B}(K^+\to\pi^+\nu\bar\nu)$ and $\mathcal{B}(K_L\to\pi^0\nu\bar\nu)$, these values for $\theta_K$ correspond to two straight lines: a horizontal one, where $\mathcal{B}(K_L\to\pi^0\nu\bar\nu)$ remains SM-like as the NP contribution is $CP$-conserving, and a slanted one which is parallel to the model-independent Grossman-Nir bound \cite{Grossman:1997sk}. This pattern is depicted by the green lines in figure \ref{fig:kpnn-branches}. If on the other hand, both left- and right-handed FCNCs are induced by NP, then neutral kaon mixing will usually be dominated by the so-called ``left-right'' effective operators containing both chiralities. In that case the correlation with the  $K\to\pi\nu\bar\nu$ decays is lost, and no correlation between  $\mathcal{B}(K^+\to\pi^+\nu\bar\nu)$ and $\mathcal{B}(K_L\to\pi^0\nu\bar\nu)$ arises. The full range for the two branching ratios, shown in red in figure \ref{fig:kpnn-branches}, is then possible. 

The discussed correlation between  $\mathcal{B}(K^+\to\pi^+\nu\bar\nu)$ and $\mathcal{B}(K_L\to\pi^0\nu\bar\nu)$ has indeed been found in a number of NP models with purely left-handed FCNC transitions \cite{Blanke:2006eb,Buras:2012dp,Buras:2012jb,Blanke:2015wba,Crivellin:2017gks}. 

\subsection{Quick Summary of Kaon Physics}

Before moving on to $B$ physics and in particular to a recent set of anomalies, let us recapitulate the unique role of kaon physics. Kaon decays have played an important role in constructing the SM, and they offfer unique opportunities to test its extensions.

{\bf The past.} In order to account for the smallness of the $K_L\to\mu^+\mu^-$ branching ratio, a fourth quark, the charm quark, has been predicted prior to its direct discovery. Also $CP$ violation has first been observed in the kaon system, by measuring a non-zero $K_L\to\pi\pi$ decay width. The necessity for a third quark generation had thus been established.

{\bf The present.} Currently, the $CP$-violating parameter $\epsilon$ places one of the most stringent constraints on physics beyond the SM, in particular if a non-trivial flavour structure is involved. In addition, the recent lattice calculations of the hadronic matrix elements entering $\varepsilon'/\varepsilon$ seem to hint for a tension between the SM prediction and the data.

{\bf The future.} If future more precise predictions of $\varepsilon'/\varepsilon$ confirm this tension, the road will be paved towards spectacular NP discoveries in rare $K$ decays. A special role is played by the $K\to\pi\nu\bar\nu$ decays, as thanks to their theoretical cleanliness they offer an extremely sensitive probe of NP.

\subsection{$B$ Physics}

Historically, kaon physics has been the main player in the field of flavour physics. More recently however, $B$ meson physics has gained significant importance. After the first observation of $B_d - \bar B_d$ oscillations, in the 1990s the two $B$-factories BaBar and Belle were build to precisely measure the properties of $B$ mesons and their decays. The $B$-factories delivered a large number of highly relevant results, including the discovery and precise measurement of $CP$ violation in $B_d - \bar B_d$ oscillations, and measurements of semileptonic $B$ decays relevant for the extraction of the CKM elements $|V_{ub}|$ and $|V_{cb}|$. Further significant improvements on the physics results of the $B$-factories, as well as measurements of a number of so far undetected rare $B$ meson decays, can be expected from the second generation $B$-factory Belle~II. First Belle~II physics results should become available within a couple of years from now.

A blind spot in the programme of the $B$-factories, however, is the physics of $B_s$ mesons. Due to their larger mass, they are not produced in the decay of the $\Upsilon(4S)$ resonance, which the $B$-factories rely on. Consequently, hadron colliders like the Tevatron and the LHC have an advantage here. Indeed, $B_s-\bar B_s$ oscillations have first been observed by the CDF experiment at Fermilab in 2006 \cite{Abulencia:2006ze}, and later confirmed by LHCb \cite{Aaij:2011qx}. The latter experiment also provides the most stringent constraint on $CP$ violation in $B_s-\bar B_s$ mixing \cite{Aaij:2014zsa}.

LHCb and, to some extend, also CMS and ATLAS have also yielded important data on a number of rare $B$ and $B_s$ meson decays, like $B\to K^{(*)}\mu^+\mu^-$ and $B_s\to\mu^+\mu^-$. For the latter, a combination of LHCb and CMS data lead to its discovery, with a branching ratio measurement in decent agreement with the SM prediction. The data on the $B\to K^{(*)}\mu^+\mu^-$ decays, on the other hand, leaves us with some intriguing anomalies to be discussed in what follows.

\subsection{\boldmath Recent Anomalies in $b \to s$ Transitions}

The aforementioned decays $B\to K\mu^+\mu^-$, $B\to K\mu^+\mu^-$, and $B_s\to\mu^+\mu^-$ are all governed by the quark level transition $\bar b\to \bar s\mu^+\mu^-$. In order to understand the physics behind the observed anomalies, we start by writing down the relevant
 effective Hamiltonian \cite{Buchalla:1995vs}
\begin{equation}\label{eq:Heff-bs}
\mathcal{H}_\text{eff}= -\frac{4 G_F}{\sqrt{2}} V_{tb} V_{ts}^{*} \frac{e^2}{16\pi^2}\sum_i(C_i {\cal O}_i +C'_i {\cal O}'_i)+h.c.\,.
\end{equation}
In the SM, only the unprimed Wilson coefficients $C_i$, corresponding to left-handed FCNC transitions, are relevant, due to the left-handedness of the flavour violating weak interactions. NP contributions, on the other hand, can have either chirality. The operators most sensitive to NP are the dipole operators 
\be
\mathcal{O}^{(\prime)}_7 =\frac{m_b}{e} (\bar s \sigma_{\mu\nu} P_{R(L)}b) F^{\mu\nu}
\ee
and the four fermion operators 
\begin{eqnarray}
\mathcal{O}^{(\prime)}_9 &=& (\bar s\gamma_\mu P_{L(R)} b)(\bar\mu\gamma^\mu\mu)\,,\\
\mathcal{O}^{(\prime)}_{10}  &=& (\bar s\gamma_\mu P_{L(R)} b)(\bar\mu\gamma^\mu\gamma_5\mu)\,,
\end{eqnarray}
that are not affected by tree-level contributions in the SM. 

The dipole operators $\mathcal{O}^{(\prime)}_7$ are constrained by the well-measured $B\to X_s\gamma$ transition, whose branching ratio is in good agreement with the SM prediction. The four-fermion operators $ \mathcal{O}^{(\prime)}_{10}$ mediate the decay $B_s\to\mu^+\mu^-$. While the data do not show a significant deviation from the SM prediction in this case, the experimental uncertainties are still sizeable, allowing for a relevant NP contribution to their Wilson coefficients $C^{(\prime)}_{10}$. The scalar and pseudoscalar four-fermion operators, on the other hand, are strongly constrained by the 
$B_s\to\mu^+\mu^-$ branching ratio measurement. We therefore neglect them in this discussion. 

The observation that different decays and observables are sensitive to different Wilson coefficients in the effective Hamiltonian \eqref{eq:Heff-bs} is crucial for the theoretical interpretation of the data. 
Of particular interest is the decay $B\to K^*\mu^+\mu^-$, where $K^*$ further decays into a kaon and a pion. The four-body final state can be described in terms of three angles and the invariant mass sqaure of the muon pair, $q^2 = (p_{\mu^+}+p_{\mu^-})^2$. The differential decay rates can then be decomposed into a sum of  contributions with specific angular dependence. We note that different parametrizations have been proposed in the literature, with the goal to minimize the theoretical uncertainties in the observables in question \cite{Altmannshofer:2008dz,Egede:2009te,Descotes-Genon:2013vna}. 

In one parametrization, a set of `optimized' observables has been derived with the goal to cancel the $B\to K^*$ form factor dependence at leading order \cite{Descotes-Genon:2013vna}. One of these observables which has attracted a lot of attention over the past few years is $P'_5$. A few years ago, the LHCb collaboration reported an anomaly in the low $q^2$ region in this observable, which is by now established at the $3.4\sigma$ level \cite{Aaij:2015oid}. Also more recent data from ATLAS \cite{ATLAS:2017dlm}
and Belle \cite{Wehle:2016yoi} hint in the same direction, albeit with much smaller significance. The recent measurement of $P'_5$ by CMS \cite{CMS-PAS-BPH-15-008}, on the other hand, is consistent with the SM. 

While the physical meaning of $P'_5$ can be understood in terms of the transversity amplitudes depending on the spin of the muon pair, its interpretation is not very intuitive and we do not go into the details here. Further information can for example be found in \cite{Descotes-Genon:2015uva}.  In what follows we focus instead on possible interpretations of the observed anomaly.

Global fits  of the Wilson coefficients in the effective Hamiltonian \eqref{eq:Heff-bs}
reveal that a sizeable non-standard contribution to $C_9$ is required to solve the $P'_5$ anomaly \cite{Capdevila:2016fhq,Beaujean:2015gba,Altmannshofer:2017fio}. Interestingly, at the same time also other, smaller tensions in the data are softened, such as the $B^+\to K^+\mu^+\mu^-$ and $B_s\to\phi\mu^+\mu^-$ branching ratios. Further, if the NP is aussmed to contribute only to the muon channel, i.\,e.\ to violate lepton flavour universality, also the $R_K$ anomaly can be explained. Here, $R_K$ is defined as the ratio of $B^+\to K^+\mu^+\mu^-$ and $B^+\to K^+ e^+e^-$ branching ratios,
\be
R_K =
\frac{\mathcal{B}(B^+\to K^+\mu^+\mu^-)}{\mathcal{B}(B^+\to K^+ e^+e^-)}\,.
\ee
The LHCb measurement of $R_K$ \cite{Aaij:2014ora} in the low $q^2$ region is 2.6 standard deviations below the very accurate SM prediction $R_K\simeq 1$ \cite{Hiller:2003js}. Similar hints for a violation of lepton flavour universality have recently also been found in the $B\to K^*\ell^+\ell^-$ ($\ell=\mu,e$) decays \cite{Wehle:2016yoi,Bifani:2017}.

The question is now how to interpret this result theoretically. Given the loop suppression of FCNCs in the SM, it is conceivable that the shift in $C_9$ is induced by NP. Popular and well-motivated NP models, such as supersymmetric theories or models with partial compositeness, can, however, not account for this deviation \cite{Altmannshofer:2013foa}. It is however possible to induce a large contribution to $C_9$ in phenomenologically viable NP models: two known examples are models with a flavour violating $Z'$ gauge boson \cite{Gauld:2013qba,Buras:2013qja,Altmannshofer:2014cfa,Chiang:2016qov,Crivellin:2016ejn}, and leptoquark scenarios \cite{Hiller:2014yaa,Bauer:2015knc,Fajfer:2015ycq}. Interestingly the latter can also adress the tension in $B\to D^{(*)}\tau\nu$ data. 

Before claiming the presence of NP in $b\to s\mu^+\mu^-$ transitions, it is however necessary to investigate the SM prediction for potentially underestimated theoretical uncertainties, see \cite{Jager:2014rwa,Altmannshofer:2017fio,Capdevila:2017ert} for recent discussions.  The main theoretical uncertainties lie, on the one hand, in the $B\to K^*$ form factors that describe the hadronic physics of the $B\to K^*$ transition in the factorization limit. On the other hand, sizeable uncertainties stem from non-factorizable corrections that arise at $\mathcal{O}(\Lambda_\text{QCD}/m_b)$. 

The hadronic form factors can be computed at large $q^2$ by lattice QCD, and at low $q^2$ by light-cone sum rule techniques. Their extrapolation yields consistent results, so that the form factors are unlikely the source of the observed anomaly. Systematic improvements of the form factor calculations can be expected over the coming years, further reducing the associated uncertainties. 

The non-factorizable corrections, however, are difficult to assess theoretically, and the associated uncertainties can only be estimated. The dominant contributions arise from long-distance charm loops coupling to photons and, in turn, the final state muon pair. In the effective Hamiltonian description of \eqref{eq:Heff-bs}, these contributions would mimic a NP contribution to the operator $\mathcal{O}_9$, due to the vector coupling of the photon. 

There are however two crucial differences that can distinguish non-factorizable corrections from NP contributions to $C_9$. First, as the photon couples universally to all lepton flavours, a lepton flavour non-universal signal would be a clear sign of NP.  If the violation of lepton flavour universality is confirmed by future data, and analogous ratios in other channels show the same pattern, then we would have an unambiguous sign of NP in semileptonic $b\to s$ transitions. Second, NP contributions are in general independent of the dimuon momentum $q^2$, while non-factorizable charm loop contributions are expected to be enhanced near the $c\bar c$ threshold. While the current data are consistent with a $q^2$-independent $C_9$, future more accurate measurements could reveal a $q^2$-dependence of the required new contribution, hence clearly disfavouring the NP interpretation. 

We have thus seen that even though the understanding of the observed $P'_5$ anomaly is currently limitied by theoretical uncertainties, future more accurate experimental data will provide a significant contribution to its resolution.

\section{Flavour Physics beyond the Standard Model}\label{sec:NP}

\subsection{The SM Flavour Problem}

As we have seen in lecture \ref{sec:SM}, flavour violation in the SM is generated by the Yukawa couplings, generating the fermion masses and flavour mixings. Most of the free parameters of the SM are related to the flavour sector, calling for a more fundamental theory that explains their origin. Moreover, the SM flavour sector is found experimentally to obey a very hierarchical pattern, with quark masses spanning five orders of magnitude, and a CKM mixing matrix close to the unit matrix. This structure seems to suggest the presence of an approximate flavour symmetry in the fundamental theory of flavour. 

Experimentally, the SM quark flavour sector has been well tested by precise measurements of a large number of flavour violating $K$, $B$, and $D$ meson decays. Despite a few anomalies, overall the SM with its simple CKM picture of flavour violation has been extremely successful at describing the data. Consequently, as we will see in section \ref{sec:constNP}, strong constraints on the scale of NP with generic flavour violating interactions can be derived. NP at the TeV scale must then have a very non-generic flavour structure, with an efficient suppression of FCNC transitions. The most widely known and employed example is the concept of Minimal Flavour Violation (MFV), which we discuss in section \ref{sec:MFV}. However it is also possible to avoid dangerously large FCNCs without imposing MFV. Different flavour symmetries and symmetry breaking patterns can be employed. A complementary approach to flavour is provided by models with partially composite fermions, where the observed flavour structure has a dynamical origin. We will briefly review this idea in section \ref{sec:comp}.

\subsection{Constraints on the Scale of New Physics}\label{sec:constNP}

In the SM, FCNC processes receive various strong suppression factors that make them highly sensitive to NP contributions. Firstly, as FCNC couplings are generated only at the loop level, they are suppressed by a loop factor $g^2/(16\pi)^2$, where $g$ is the weak $SU(2)_L$ coupling constant. The GIM mechanism further reduces the size of FCNC transitions, in particular in the kaon system. FCNC transitions in the $K$, $B_d$ and $B_s$ systems, respectively, are then governed by the following CKM factors:
\be\label{eq:CKMsup}
\underbrace{V_{ts}^* V_{td}}_{K\text{ system}} \sim 5\cdot 10^{-4} \,, \qquad 
\underbrace{V_{tb}^* V_{td}}_{B_d\text{ system}} \sim  10^{-2} \,, \qquad
\underbrace{V_{tb}^* V_{ts}}_{B_s\text{ system}} \sim  4\cdot 10^{-2} \,.
\ee
We observe that the CKM hierarchy yields the strongest suppression in the kaon system, while $b\to d$ and in particular $b\to s$ transitions are much less rare. Lastly, due to the left-handedness of weak interactions, FCNC processes in the SM are purely left-handed. As we will see below, the purely left-handed effective operators are much less affected by renormalization group effects than the left-right ones that are generated in many NP models. 

All of these suppression mechanisms can in principle be circumvented by NP, so that FCNC transitions provide an excellent sensitivity to NP even much beyond the TeV scale. To explore the NP reach of flavour physics in a model-independent way, it is useful to study it in terms of the effective field theory language. In this framework, the renormalizable SM Lagrangian is extended by including all higher-dimensional effective operators that are consistent with the gauge symmetries of the SM:
\be\label{eq:LEFT}
\mathcal{L}_\text{EFT} = \mathcal{L}_\text{SM} + \sum_i \frac{C_i}{\Lambda}\mathcal{O}^\text{dim 5}_i+ \sum_i \frac{C_i}{\Lambda^{2}}\mathcal{O}^\text{dim 6}_i+\cdots\,.
\ee
The SM then constitutes the low energy limit valid at energy scales much below $\Lambda$, where the higher-dimensional operator contributions are irrelevant. The scale $\Lambda$ is the cut-off scale of the effective theory. It generally arises from integrating out new particles with masses $M\sim\mathcal{O}(\Lambda)$. Hence, at energy scales above $\Lambda$ it has to be replaced by the full theory in which the new particles are physical degrees of freedom. 

The only operator of dimension five in \eqref{eq:LEFT} is the Weinberg operator \cite{Weinberg:1979sa} that is relevant for the generation of neutrino masses. For details, see the lectures of Gabriela Barenboim \cite{Barenboim:2016ili}. The leading operators mediating FCNCs arise at dimension six and are therefore suppressed by two  powers of the inverse of the cut-off scale $\Lambda$. 

In a general NP scenario, FCNC transitions are not related to the CKM matrix and the respective CKM suppression factors in \eqref{eq:CKMsup} are absent. We can instead parametrize the strength of flavour violating transitions by a parameter $\delta$ that can in general depend on the meson system in question. Then the operators contributing to neutral meson mixings ($\Delta F = 2$) are proportional to $\delta^2$, while the operators contributing to rare decays like $K\to\pi\nu\bar\nu$ violate flavour only by one unit ($\Delta F = 1$) and are therefore proportional to $\delta$. Hence if flavour violation is suppressed in the NP sector, i.\,e.\ $\delta \ll 1$, then rare decays are in general more sensitive to the contributions from NP than $\Delta F =2$ observables. If, on the other hand, FCNC effects are suppressed by a large NP scale, $\Lambda \gg M_W$ and $\delta\sim\mathcal{O}(1)$, then $\Delta F = 2$ observables are typically more sensitive to NP effects than $\Delta F =1$ ones. The sensitivity to large NP scales increases with increasing flavour violation $\delta$ in the NP sector, and for sizeable values of $\delta$ extends far beyond the reach of the LHC.

To investigate the NP reach of flavour physics, specifically of $\Delta F=2$ transitions, more explicitly, let us consider the general dimension six effective Hamiltonian:
\be\label{eq:Heff}
\mathcal{H}_\text{eff}^{\Delta F = 2} = \frac{1}{\Lambda^2}\left[\sum_{i=1}^5 C_i \mathcal{O}_i + \sum_{i+1}^3 \tilde C_i \tilde{\mathcal{O}}_i\right]\,.
\ee
Here, the four fermion operators mediating $B_{d,s}-\bar B_{d,s}$ mixing are defined as ($q=d,s$)
\begin{align}
& \mathcal{O}_1 = (\bar q^\alpha \gamma_\mu P_L b^\alpha)(\bar q^\beta \gamma^\mu P_L b^\beta) \,, &&  \\
& \mathcal{O}_2  = (\bar q^\alpha  P_L b^\alpha)(\bar q^\beta P_L b^\beta) \,,&& 
\mathcal{O}_3  = (\bar q^\alpha  P_L b^\beta)(\bar q^\beta P_L b^\alpha) \,,\\
&\mathcal{O}_4  = (\bar q^\alpha  P_L b^\alpha)(\bar q^\beta P_R b^\beta) \,,&& 
\mathcal{O}_5  = (\bar q^\alpha  P_L b^\beta)(\bar q^\beta P_R b^\alpha) \,,\\
& \tilde{\mathcal{O}}_1 = (\bar q^\alpha \gamma_\mu P_R b^\alpha)(\bar q^\beta \gamma^\mu P_R b^\beta) \,, &&  \\
& \tilde{\mathcal{O}}_2  = (\bar q^\alpha  P_R b^\alpha)(\bar q^\beta P_R b^\beta) \,, && 
 \tilde{\mathcal{O}}_3  = (\bar q^\alpha  P_R b^\beta)(\bar q^\beta P_R b^\alpha) \,,
\end{align}
where summation over the colour indices $\alpha,\beta$ is understood. Analogous expressions hold for the operators mediating $K^0-\bar K^0$ and $D^0-\bar D^0$ mixing. In the SM, ony the operator $\mathcal{O}_1$ is present.

\begin{figure}
\center{\includegraphics[width=.4\textwidth]{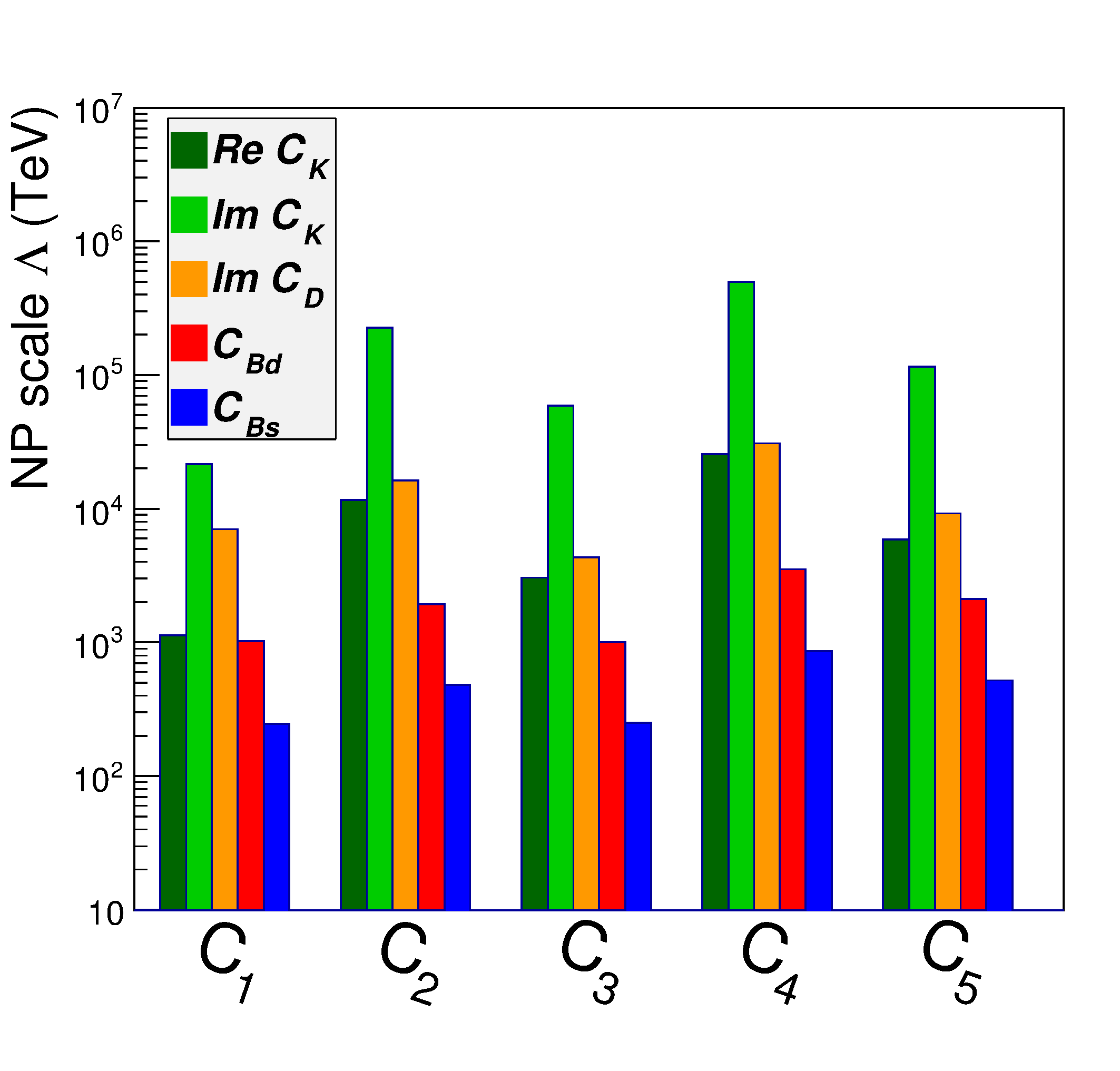}}
\caption{Model-independent constraints on the NP scale $\Lambda$, obtained from the various operators of the $\Delta F =2$ effective Hamiltonian. Figure taken from \cite{UTfit-ICHEP16} with kind permission of the authors.\label{fig:Lambda}}
\end{figure}

Assuming a generic NP flavour structure, i.\,e.
\be
|C_i|\sim \mathcal{O}(1)\,, \qquad \arg C_i \sim\mathcal{O}(1)\,,
\ee
it is possible to derive lower bounds on the NP scale by comparing the data on meson mixing observables with their theory predictions \cite{UTfit-ICHEP16}. The result of this exercise is shown in figure \ref{fig:Lambda}. The constraints from different meson systems are shown in dfferent colours. We find that, as expected, the strongest constraints arise from neutral kaon mixing, and in particular from the $CP$-violating parameter $\varepsilon$. Most constraining are the operators $\mathcal{O}_2$ and $\mathcal{O}_4$, which constrain the scale of generic NP to be above $10^5\,\text{TeV}$. Stringent bounds are also obtained from the non-observation of $CP$-violation in neutral $D$ meson mixing. The $B_d$ and $B_s$ systems, although less constraining, still push the NP scale above $100\,\text{TeV}$.

We conclude that NP at or near the TeV scale, required for a natural origin of EW symmetry breaking, mut have a very non-trivial flavour structure. Clearly, flavour can not be a conserved quantum number in the NP model, as the flavour symmetry is already broken in the SM. Yet it is possible to achieve an approximate conservation of flavour, as we will see in what follows. 

\subsection{Minimal Flavour Violation}\label{sec:MFV}

In lecture \ref{sec:SM} we have seen that the SM quark sector possesses a global flavour symmetry 
\be
G_\text{flavour} = U(3)_Q \times U(3)_U \times U(3)_D
\ee
that is explicitly broken by the quark Yukawa couplings $Y_U$ and $Y_D$. Due to the hierarchical structure of the Yukawa matrices, FCNC processes in the SM are strongly suppressed. The idea is then to extend this mechanism to the NP sector in order to suppress flavour violation also beyond the SM and reconcile TeV scale NP with the stringent experimental flavour constraints.

This leads us to formulate to the {\it Minimal Flavour Violation (MFV)} hypothesis 
\cite{Buras:2000dm,Buras:2003jf,DAmbrosio:2002vsn}: In MFV,  also in the NP sector the quark Yukawa couplings $Y_U$ and $Y_D$ consitute the only source of breaking of the  flavour symmetry $G_\text{flavour}$. For phenomenological reasons, it is usually also assumed that no new $CP$-violating phases arise, so that the only source of $CP$ violation remains the CKM phase. The MFV hypothesis ensures that all flavour and $CP$-violating NP effects are suppressed by the same CKM factors as the corresponding SM ones. 

To impose the MFV hypothesis, it is useful to think of the Yukawa couplings as so-called spurion fields. That means, we can formally restore the flavour symmetry $G_\text{flavour}$ by treating the Yukawa couplings $Y_U, Y_D$ as dimensionless auxiliary (i.\,e.\ non-dynamical) fields. Assigning the  $G_\text{flavour}$ transformation properties
\be
Y_U \sim (3,\bar 3, 1)\,, \qquad Y_D \sim (3,1,\bar 3)
\ee
restores the flavour symmetry of the SM Lagrangian. In particular, the Yukawa coupling Lagrangian \eqref{eq:Yuk} becomes (formally) invariant under the flavour symmetry. 

In order to extend this concept to the NP sector, we consider again the effective Lagrangian in \eqref{eq:LEFT}. The MFV ansatz requires us to render the higher-dimensional effective NP operators invariant, by expanding the Wilson coefficients $C_i$ in terms of appropriate combinations of $Y_U$, $Y_D$.

As we are mainly interested in FCNC processes in the down quark sector, it is convenient to work in the down quark mass basis. In this basis, the Yukawa matrices simplify to
\be
Y_D = \text{diag}\,(y_d,y_s,y_b)\,, \qquad Y_U = \hat V_\text{CKM}^\dagger\, \text{diag}\,(y_u,y_c,y_t) \,,
\ee 
where $y_i = m_i/v$.

Let us now consider the operator $\mathcal{O}_1$ in the $\Delta F =2$ effective Hamiltonian \eqref{eq:Heff}:
\be
\frac{C_1^{ij}}{\Lambda^2}(\bar Q_{Li} \gamma_\mu Q_{Lj}) (\bar Q_{Li}\gamma^\mu Q_{Lj})\,,
\ee
where $i\ne j $ are flavour indices. Restoring the flavour symmetry $G_\text{flavour}$, we find the following expression for the Wilson coefficient $C_1^{ij}$ in terms of the Yukawa couplings $Y_U,Y_D$:
\begin{eqnarray}
C^{ij}_1 &=& (a \cdot \mathbbm{1} + b\cdot Y_D Y_D^\dagger + c \cdot Y_U Y_U^\dagger+\cdots)_{ij}^2\nonumber\\
&=& (c\cdot \hat V_\text{CKM}^\dagger \,\text{diag}\, (y_u^2,y_c^2,y_t^2) \hat V_\text{CKM})_{ij}^2\nonumber\\
&\simeq& c^2\cdot y_t^4 (V_{ti}^* V_{tj})^2 \,.\label{eq:O1}
\end{eqnarray}
Here $a,b,c$ are real expansion parameters that are assumed to be $\mathcal{O}(1)$.
We confirm that the NP contribution to $C_1$ is  suppressed by the same CKM factors \eqref{eq:CKMsup} as the SM contribution. 

Next, let us have a look at the operator $\mathcal{O}_4$,
\be
 \frac{C_4^{ij}}{\Lambda^2}(\bar D_{Ri} Q_{Lj})(\bar Q_{Li} D_{Rj})\,,
\ee
which was found to generate the strongest constraints on the scale of generic NP contributions. Again using the MFV hypothesis, we find
\begin{eqnarray}
C_4^{ij} &=& \left[Y_D^\dagger (a\cdot\mathbbm{1}+ b\cdot Y_D Y_D^\dagger + c\cdot Y_U Y_U^\dagger+\cdots)\right]_{ij} \nonumber\\
&& \times \left[(d\cdot\mathbbm{1}+ e\cdot Y_D Y_D^\dagger  + f\cdot Y_U Y_U^\dagger +\cdots) Y_D\right]_{ij} \nonumber\\
&=& c\cdot {y_i} y_t^2 (V_{ti}^* V_{tj}) \times f\cdot { y_j} y_t^2 (V_{ti}^* V_{tj})\,.
\end{eqnarray}
Again $a,b,c,d,e,f$ are real and $\mathcal{O}(1)$ expansion parameters. We observe that in MFV the Wilson coefficient $C_4$ is strongly suppressed not only by CKM elements, but in addition also by the masses of the external quarks $i$ and $j$. The stringent constraints can therefore be evaded.

This observation is actually quite general. Whenever a right-handed down quark appears in a higher-dimensional operator, in the MFV framework it must necessarily be accompanied by at least one power of the Yukawa coupling $Y_D$. As the top mass is found to be much larger than the bottom mass, in the SM as well as in many concrete NP models, the hierarchy $y_t\gg y_b$ holds. In this case, operators that involve additional $Y_D$ factors become negligible. Therefore, only those operators are relevant for $K$ and $B$ physics that involve flavour transitions of left-handed quark fields $Q_L$. These are the ones that are already present in the SM effective Hamiltonian, due to the left-handedness of the weak interactions. Note however that not in all NP models the relation $y_t\gg y_b$ holds. It can be violated in models with extended Higgs sectors, such as the MSSM at large $\tan\beta$.

These considerations lead us to the definition of a slightly more restrictive version of MFV, the framework of {\it Constrained Minimal Flavour Violation (CMFV)} \cite{Buras:2000dm,Buras:2003jf,Blanke:2006ig}. Again, in CMFV the global quark flavour symmetry $G_\text{flavour}$ is broken only by the SM Yukawa couplings $Y_U$ and $Y_D$, and no new $CP$-violating phases are present. In addition to these MFV assumptions, in CMFV only those effective operators are relevant that are present already in the SM.

In the $\Delta F =2$ effective Hamiltonian then only the operator $\mathcal{O}_1$ remains. As shown in \eqref{eq:O1}, in MFV models it has the same CKM dependence $(V_{ti}^* V_{tj})^2$ as the SM contribution. Consequently, in the CMFV scenario we can parametrise the NP contributions to meson mixing observables by a real and flavour-universal shift in the loop function
\be
S_0(x_t) \to S(p)\,,
\ee
where $p$ collectively denotes the parameters of a given CMFV model. Hence in CMFV models the $\Delta F = 2$ sector can be described by a single new parameter, and the NP contributions to neutral meson mixings in the various meson systems are correlated. 

We can make use of this feature to construct the unitarity triangle in figure \ref{fig:UT} from $\Delta F = 2 $ observables in such a way that the result holds for all CMFV models, independent of the value of the function $S(p)$ \cite{Buras:2000dm}. As always, the value of $|V_{us}|$ has to be determined from tree level decays, and is precisely known. The angle $\beta$ can be measured in the time-dependent $CP$ asymmetry of the $B_d\to J/\psi K_S$ that measures $CP$ violation in $B_d-\bar B_d$ mixing. Finally, the length of the side $R_t$ is proportional to the square root of the ratio $\Delta M_d/\Delta M_s$, where the NP contribution cancels out in the ratio. 

The thus determined {\it universal unitarity triangle} is currently much more precisely known \cite{Blanke:2016bhf} than the one determined solely from tree level decays. While the two determinations currently show a good agreement, a potential mismatch observed with future more precise tree level data, would be an unambiguous sign of physics beyond the CMFV hypothesis. 

We have thus seen that the MFV ansatz allows us to lower the scale of NP to the TeV range, without inducing large NP contributions to FCNC observables. In particular the CMFV hypothesis provides a very predictive framework, inducing many correlations between FCNC observables. A pedagogical review can be found in \cite{Buras:2003jf}. However, in the next section we will see that MFV is not the only option to reconcile TeV-scale NP with the flavour data.

\subsection{Flavour Hierarchies from Partial Compositeness}\label{sec:comp}

It is also possible to suppress flavour violating processes without introducing flavour symmetries. Models with partially composite fermions are a well-known example where flavour hierarchies have a dynamical origin. In this section we only provide a brief and superficial overview on the flavour structure of composite models. More detailed, excellent lectures on composite Higgs models and their dual 5D  description in the Randall-Sundrum framework can be found e.\,g.\ in \cite{Csaki:2004ay,Sundrum:2005jf,Csaki:2005vy,Contino:2010rs,Gherghetta:2010cj,Ponton:2012bi}.

The basic idea of composite Higgs models is the realization of the Higgs boson as a light composite state of a strongly coupled sector in analogy to the pions of QCD. The naturalness problem of fundamental scalars is thereby avoided. The EW symmetry is assumed to be broken by a condensate of the composite sector. 
Further composite resonances are then expected at the TeV scale, similar to the $\rho, \omega\, \dots$ mesons of QCD with masses an order of magnitude larger than the pion mass.  In order to avoid constraints from precision tests of the SM, the SM particles -- except for the Higgs boson and the top quark -- have to be mostly elementary.  

\begin{figure}
\center{\includegraphics[width=.28\textwidth]{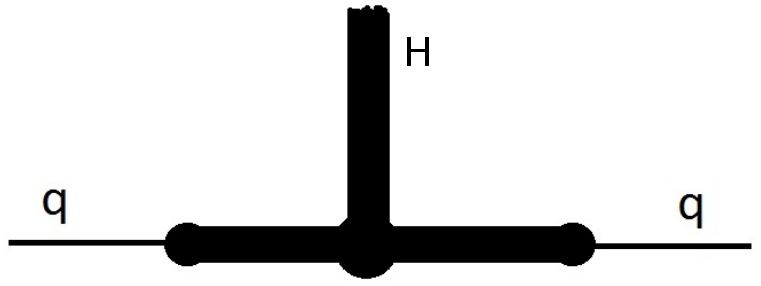}}
\caption{Effective Yukawa coupling in models with partially composite fermions.}\label{fig:compY}
\end{figure}

In traditional composite models, like Technicolour \cite{Kaul:1981uk,Lane:2002wv}, the generation of realistic fermion masses turned out to be a major problem.
As a solution, more recently the concept of partially composite fermions has been put forward \cite{Kaplan:1991dc}. In this setup, the known SM fermions and gauge bosons form an elementary sector. The composite sector, responsible for EW symmetry breaking, gives rise to operators describing conposite resonances that have quantum numbers identical to the ones of the SM fields. The elementary fermions $Q, U, D$ of \eqref{eq:quarks} are coupled to the composite sector by a linear mixing with the composite operators $\mathcal{O}_Q, \mathcal{O}_U, \mathcal{O}_D$:
\be
\mathcal{L}_\text{mixing} = \epsilon_Q\bar Q_L \mathcal{O}_Q + \epsilon_U\bar U_R \mathcal{O}_U + \epsilon_D \bar D_R \mathcal{O}_D\,.
\ee
The observed Yukawa couplings $Y_U,Y_D$ are then a combination of the strong sector coupling responsible for the interactions $\lambda_U,\lambda_D$ among the composite resonances and the elementary-composite fermion mixings $\epsilon_Q,\epsilon_U,\epsilon_D$,
\be\label{eq:Yeff}
Y_{U,D} = \epsilon_Q \,\epsilon_{U,D}\, \lambda_{U,D}\,.
\ee
Figure \ref{fig:compY} displays the generation of the effective Yukawa couplings diagrammatically. 
As the strong sector couplings are in general expected to be structureless (``anarchic''), 
\be
|\lambda_{U,D}^{ij}| \sim\mathcal{O}(1) \qquad \arg \lambda_{U,D}^{ij} \sim\mathcal{O}(1) \,,
\ee
the flavour hierarchies in the effective SM Yukawa couplings has to be induced by hierarchies in the elementary-composite mixing. 

The stringent experimental constraints on non-SM interactions of the first two quark generations tell us that the latter must be mostly elementary, i.\,e.\ their mixing with the composite sector is small:
\be
\epsilon_{Q}^{1,2},\epsilon_{U}^{1,2},\epsilon_{D}^{1,2}\ll 1\,.
\ee
The third generation is less constrained, and indeed the large top quark Yukawa coupling requires that
\be
\epsilon_Q^3 \sim \epsilon_U^3 \sim\mathcal{O}(1)\,.
\ee
Overall, the observed quark masses and the experimental precision constraints imply the pattern
\be
\epsilon_Q^1 \ll \epsilon_Q^2 \ll \epsilon_Q^3 \sim \mathcal{O}(1) \,,\qquad 
\epsilon_U^1 \ll \epsilon_U^2 \ll \epsilon_U^3 \sim \mathcal{O}(1) \,,\qquad
\epsilon_D^1 \ll \epsilon_D^2 \ll \epsilon_D^3 \ll 1\,.
\ee

Inserting this pattern into  \eqref{eq:Yeff} and assuming an anarchic composite sector, we find the following hierarchical structure for the observed Yukawa couplings
\be
Y_U\sim
\begin{pmatrix}
\epsilon_Q^1 \epsilon_U^1 & \epsilon_Q^1 \epsilon_U^2& \epsilon_Q^1 \epsilon_U^3 \\
\epsilon_Q^2 \epsilon_U^1 & \epsilon_Q^2 \epsilon_U^2& \epsilon_Q^2 \epsilon_U^3 \\
\epsilon_Q^3 \epsilon_U^1 & \epsilon_Q^3 \epsilon_U^2& \epsilon_Q^3 \epsilon_U^3 
\end{pmatrix}\,,
\qquad
Y_D\sim
\begin{pmatrix}
\epsilon_Q^1 \epsilon_D^1 & \epsilon_Q^1 \epsilon_D^2& \epsilon_Q^1 \epsilon_D^3 \\
\epsilon_Q^2 \epsilon_D^1 & \epsilon_Q^2 \epsilon_D^2& \epsilon_Q^2 \epsilon_D^3 \\
\epsilon_Q^3 \epsilon_D^1 & \epsilon_Q^3 \epsilon_D^2& \epsilon_Q^3 \epsilon_D^3 
\end{pmatrix}\,.
\ee
Note that the hierarchical structure of the effective Yukawa couplings $Y_U,Y_D$ is analogous to the one obtained in models with a  Froggatt-Nielsen \cite{Froggatt:1978nt} flavour symmetry \cite{Casagrande:2008hr,Blanke:2008zb,Albrecht:2009xr}.

Diagonalizing these matrices, we not only recover the observed quark mass hierarchy, but we also find predictions for the off-diagonal elements of the CKM matrix \cite{Huber:2000ie,Huber:2003tu}:
\be
|V_{us}| \sim \frac{\epsilon_Q^1}{\epsilon_Q^2} \ll 1\,,\qquad 
|V_{cb}| \sim \frac{\epsilon_Q^2}{\epsilon_Q^3} \ll 1\,,\qquad
|V_{ub}| \sim \frac{\epsilon_Q^1}{\epsilon_Q^3} \ll 1\,.
\ee
The measurements of $|V_{us}|$ and $|V_{cb}|$ fix the hierarchies among the $\epsilon_Q^i$ parameters. We then obtain a prediction for $|V_{ub}|$:
\be\label{eq:Vub-pred}
 |V_{ub}| \sim |V_{us}| \cdot |V_{cb}| \sim 0.2\,\cdot\, 4\cdot10^{-2} =8\cdot 10^{-2}\,.
\ee 
This number is larger than the measured value of $|V_{ub}|$ by a factor of two. Keeping in mind that in the derivation of \eqref{eq:Vub-pred} we dropped $\mathcal{O}(1)$ factors, this result is quite remarkable. 

\begin{figure}[h]
\center{\includegraphics[width=.28\textwidth]{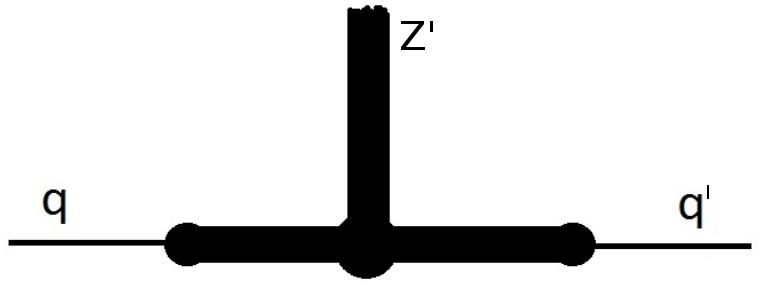}}
\caption{Tree level flavour changing coupling of a composite $Z'$ gauge boson in models with partially composite fermions.}\label{fig:compZp}
\end{figure}

In addition to providing a dynamical origin for the observed pattern of the SM quark masses and CKM mixings, the hierarchies in the elementary-composite fermion mixing also efficiently suppress tree level FCNC couplings \cite{Agashe:2004cp} that are generated in the composite sector and mediated to the SM fermions by the elementary-composite mixing. Assuming again that the composite sector couplings are anarchic, the FCNC couplings of the SM quarks are suppressed by the same hierarchical pattern as the effective Yukawa couplings $Y_U,Y_D$, as can be seen from comparing the FCNC coupling in figure \ref{fig:compZp} to the Yukawa coupling in figure \ref{fig:compY}.

With the masses of the composite resonances at a few TeV, as required by EW precision data and direct LHC searches, the suppression of FCNC couplings is sufficient to comply with most of the available flavour data. Some tension arises however with $CP$-violating observables in the kaon sector, specifically $\varepsilon$ \cite{Csaki:2008zd} and $\varepsilon'$ \cite{Gedalia:2009ws}. The latter constraints require the masses of the strong sector resonances to be above 10\,TeV, unless some additional structure is imposed on the composite sector to evade the constraints. 

In models with partially composite fermions, the SM flavour hierarchies is thus traced back to the exponentially small mixing of elementary fermions with the composite sector containing the Higgs. 
But where does this exponential suppression come from? 

The origin of the flavour hierarchies in partially composite models can be best understood by considering the {\it holographic dual} of these theories. In classical optics, holograms  are known as 2D images containing the information of a 3D object. In that sense, the 3D object and its 2D holographic image are {dual} to each other: they contain the same information. In a similar way, it has been proposed that a 4D composite model is, under certain conditions, dual to a 5D weakly coupled model. By studying the properties of the dual 5D model, we can then obtain a better understanding of the 4D composite model.

\begin{figure}[h]
\center{\includegraphics[width=.33\textwidth]{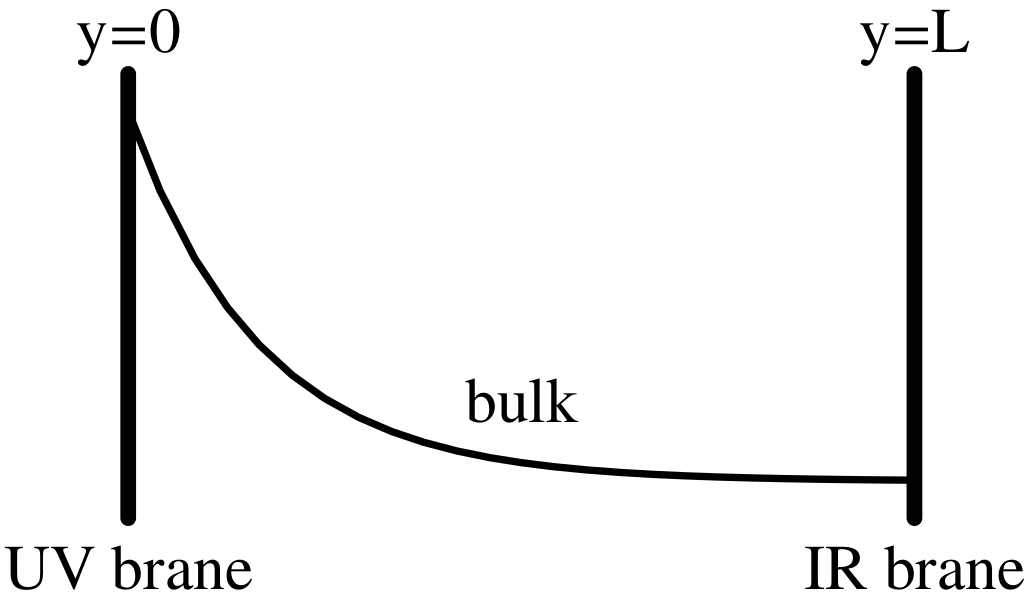}}
\caption{RS space-time.\label{fig:RS}}
\end{figure}

The foundation for the construction of the 5D holographic dual is laid by the AdS/CFT-corres\-pondence \cite{Maldacena:1997re}. It has been conjectured that a strongly coupled 4D conformal theory is dual to a weakly coupled 5D theory in the Randall-Sundrum (RS) \cite{Randall:1999ee} background. Conformal symmetry is a particular internal symmetry of the strongly coupled sector -- for our purposes it is sufficient to know that conformality implies scale invariance. The same symmetry group, $SO(4,2)$ can also be implemented as a space-time symmetry. In order to achieve this, our 4D space-time has to be extended by one extra dimension. The symmetry of the 5D space-time is then described by the RS metric \cite{Randall:1999ee}
\be\label{eq:RS}
ds^2 = {e^{-2ky}}\eta_{\mu\nu}dx^\mu dx^\nu - dy^2\,,
\ee
displayed in figure \ref{fig:RS}.
The extra dimensional coordinate $y$ is confined to an interval $0\le y \le L$. In the dual strongly coupled theory, the coordinate $y$ corresponds to the energy scale $\Lambda$ of the theory. The endpoint $y=0$, called UV brane, corresponds to the Planck scale where the conformal symmetry is explicitly broken. The other endpoint $y=L$, the IR brane, corresponds to the TeV scale, where a strong sector condensate spontaneously breaks conformality.

The change of energy scales along the fifth coordinate $y$ can be understood by having another look at the RS metric \eqref{eq:RS}. The ``warp factor'' ${e^{-2ky}}$ induces a $y$ dependence of the effective energy scale 
\be
\Lambda(y) = e^{-ky}\Lambda_0
\ee
in terms of the fundamental energy scale $\Lambda_0$ of the theory. The large hierarchy between the Planck scale and the scale of EW symmetry breaking can then be explained by localizing the Higgs boson on the IR brane where the effective cut-off scale is warped down to the TeV scale.

Due to the confinement of the coordinate $y$ to an interval, the fifth dimension is too small to be directly observed at low energies, and our world effectively appears four-dimensional. However the extra dimension leaves observable traces in terms of its 4D remnants, the Kaluza-Klein (KK) modes \cite{Gherghetta:2000qt}. To understand their origin, let us recall the description of a potential well in quantum mechanics. Due to the wave function confinement to an interval, an infinite tower of discrete modes appears, with quantized energy levels.
The same concept applies when considering 5D fields, when the fifth dimension is confined to an interval: Integrating out the dynamics of the extra dimension, an infinite tower of 4D fields arises, with identical quantum numbers and increasing masses. The lowest-lying mode is massless and identified with the corresponding SM field. The lowest excited modes have masses in the TeV range. They are the dual states of the composite resonances in the strongly coupled 4D theory.

\begin{figure}[h]
\center{\includegraphics[width=.4\textwidth]{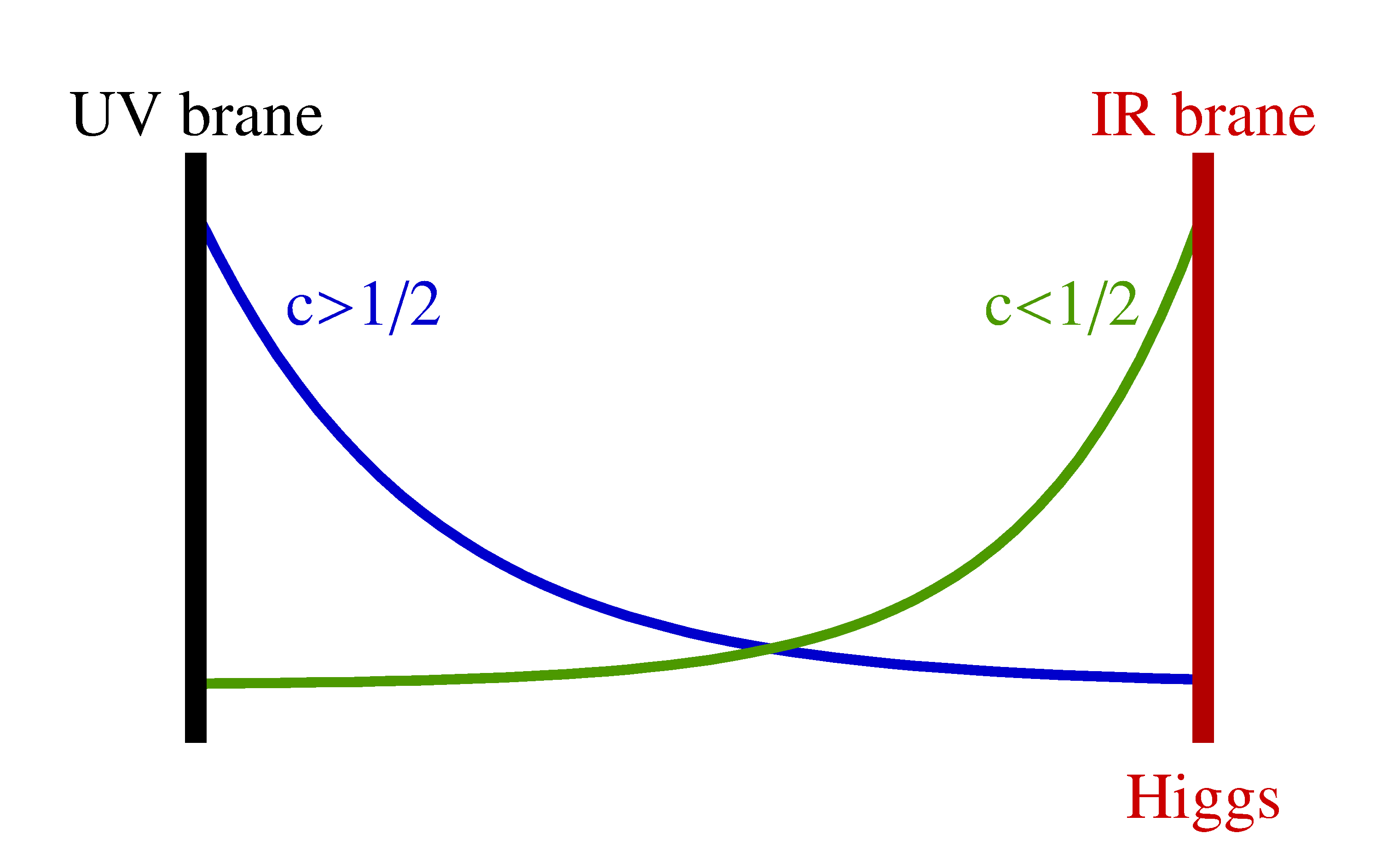}}
\caption{Fermion zero mode localisation in the RS background.\label{fig:RSferm}}
\end{figure}

Of particular interest is the fermion sector of RS models. The localization of the fermionic zero modes, corresponding to the SM fermions, is described by the wave function \cite{Grossman:1999ra,Gherghetta:2000qt,Burdman:2002gr}
\be
f^{(0)}(y,c) \propto {e^{(\frac{1}{2}-{c})ky}}\,,
\ee
where $c$ is the bulk mass parameter of the respective fermion, a fundamental parameter of the 5D Lagrangian. As for all dimensionless parameters, naturally $c\sim\mathcal{O}(1)$.  The localization of a given fermion zero mode and its overlap with the Higgs boson wave function localized on the IR brane hence exponentially depends on its bulk mass parameter $c$, as shown in figure \ref{fig:RSferm}. 

The effective 4D Yukawa couplings then have the following structure:
\be \label{eq:RS-Y}
{(Y_{U,D})_{ij}} \sim  {{f^Q_i\, f^{U,D}_j}}\, {{(\lambda_{U,D})_{ij}}}\,.
\ee
Here, $\lambda_{U,D}$ are the fundamental Yukawa couplings of the full 5D theory that are assumed to be anarchic, and $f^Q_i, f^{U,D}_j$ are the relevant fermion zero mode wave functions evaluated on the IR brane. 

Comparing \eqref{eq:RS-Y} to the effective Yukawa couplings in partially composite fermion models, \eqref{eq:Yeff}, we can straightforwardly deduce the dual interpretation of the fermion zero mode localization. Fermion zero modes with $c< 1/2$ are localized close to the IR brane and couple strongly to the Higgs boson. These are the heavy fermions that in the 4D theory have a large elementary-composite mixing. The light, mostly elementary fermions, on the other hand, have $c> 1/2$ and are therefore localized near the UV brane. Consequently their coupling to the Higgs boson is exponentially suppressed. We  conclude that the localization of fields along the 5D bulk corresponds to their degree of compositeness in the 4D dual theory. 

Making use of the AdS/CFT correspondence, we have thus found a dynamical origin for the exponential suppression of the elementary-composite mixing of the light fermions in partially composite models. In the 5D dual of these theories, the exponential flavour hierarchies arise from different values for their bulk mass parameters, all naturally of $\mathcal{O}(1)$. These models therefore provide an appealing alternative to models with approximate flavour symmetries. The flavour phenomenology of RS models with bulk fermions has been the subject of many detailed studies, see e.\,g.\ \cite{Agashe:2004ay,Csaki:2008zd,Blanke:2008zb,Blanke:2008yr,Bauer:2009cf,Blanke:2012tv}.

\section*{Summary and Outlook }

This lecture series provided a basic introduction to flavour physics and $CP$ violation, as well as an overview over some current hot topics.  

In lecture \ref{sec:SM} we reviewed the basics of flavour physics in the SM. The flavour symmetry of the SM is violated by the Yukawa couplings, which give rise to the quark masses and the CKM mixing matrix. Their observed very hierarchical structure constitutes the {SM flavour problem} and calls for a more fundamental theory of flavour. We also outlined the theoretical description of flavour violating processes in the SM and beyond, using the effective Hamiltonian and the operator product expansion. 

Lecture \ref{sec:pheno} was devoted to the discussion of some basic phenomenological concepts in $K$ and $B$ meson physics. As flavour physics in the quark sector has a very rich phenomenology, we restricted our attention to a few but very important representative examples here. We introduced the physics of neutral kaon mixing and $CP$ violation in $K\to\pi\pi$ decays, described by the parameters $\varepsilon$ and $\varepsilon'$. We then turned our attention to the very rare decays $K^+\to\pi^+\nu\bar\nu$ and $K_L\to\pi^0\nu\bar\nu$, which, due to their theoretical cleanliness, offer an excellent probe of NP even at large energy scales. In the $B$ system, we focused on the semileptonic $b\to s\mu^+\mu^-$ transitions, for which several anomalies have been found in recent data. 

In lecture \ref{sec:NP} we gave an introduction to flavour physics in theories beyond the SM. We started by identifying the stringent constraints on the scale of generic NP contributions obtained from the neutral meson mixing observables. We then introduced two concepts that suppress large non-standard contributions to FCNC observables and thereby reconcile TeV-scale NP with flavour data. In Minimal Flavour Violation, the SM Yukawa couplings are assumed to be the only source of flavour and $CP$ violation also in the NP sector. Consequently, the new flavour violating effects are suppressed by the same hierarchical structures as in the SM, and a predictive pattern of correlations arises. A dynamical origin of flavour hierarchies, on the other hand, is provided by models with partially composite fermions, which are dual to 5D theories in the Randall-Sundrum background. 

Flavour physics has played an essential role in the construction of the SM. More recently, its importance has shifted from measuring the parameters of the SM to hunting for possible NP contributions. With the lack of direct discoveries of new particles at the LHC, indirect searches for NP are becoming increasingly relevant. Indeed, some of the most convincing anomalies of today's particle physics are related to the flavour sector. In addition, over the coming years, a lot of experimental progress will be made, with the potential for striking NP discoveries.

\section*{Acknowledgements}

I am grateful to the organizers of the ESHEP 2016 school for inviting me to give the series of lectures presented here, and for putting so much time and effort into making this school a very successful and enjoyable one. I also thank all participants for the pleasant and fruitful atmosphere in Skeikampen. Last but not least I am much obliged to the Jotunheimen mountain guides for not leaving me behind on the Besseggen ridge.

\end{document}